\documentclass[11pt]{article}
\usepackage[left=1in,right=1in,top=1in,bottom=1in]{geometry} 
\usepackage{amsmath, amssymb, amsthm} 
\usepackage{graphicx,psfrag} 
\usepackage{chngcntr} 
\usepackage{caption} 
\usepackage{subcaption}
\usepackage{mathtools} 
\usepackage[mathscr]{euscript} 
\usepackage{bm, bbm} 
\usepackage{enumerate}
\usepackage{natbib} 
\usepackage{url} 
\usepackage{lipsum}
\usepackage[bottom]{footmisc} 
\usepackage{booktabs} 
\usepackage{varwidth}
\usepackage{multirow} 
\usepackage{makecell}
\usepackage{tabularx}
\newcolumntype{Y}{>{\centering\arraybackslash}X}
\usepackage[ruled,vlined]{algorithm2e}
\usepackage{xcolor} 
\usepackage[hypertexnames=false]{hyperref} 
\hypersetup{
colorlinks = true,
allcolors = blue 
}
\usepackage{cleveref} 
\usepackage{pdflscape}
\usepackage{minitoc}   
\usepackage{authblk} 
\usepackage{setspace}
\makeatletter
\newcommand{\algorithmfootnote}[2][\footnotesize]{%
  \let\old@algocf@finish\@algocf@finish
  \def\@algocf@finish{\old@algocf@finish
    \leavevmode\rlap{\begin{minipage}{\linewidth}
    #1#2
    \end{minipage}}%
  }%
}
\makeatother

\usepackage{macros} 

\begin{document}

\title{A scalable two-stage Bayesian approach accounting for exposure measurement error in environmental epidemiology} 

\author{CHANGWOO J. LEE\\ {\vspace{-4mm} \small \it Department of Statistics, Texas A\&M University, College Station, TX 77843, USA} \and
        \vspace{-2mm} ELAINE SYMANSKI\\ {\vspace{-4mm} \small \it Center for Precision Environmental Health, Departments of Medicine and Family and Community Medicine, Baylor College of Medicine,  Houston, TX 77030, USA}
        \and 
        \vspace{-2mm} AMAL RAMMAH\\ {\vspace{-4mm} \small \it Center for Precision Environmental Health, Baylor College of Medicine,  Houston, TX 77030, USA}
        \and 
        \vspace{-2mm} DONG HUN KANG \\ {\vspace{-4mm} \small \it Texas A\&M Transportation Institute, Texas A\&M University System, College Station, TX 77843, USA} 
        \and
        \vspace{-2mm} PHILIP K. HOPKE\\ {\vspace{-4mm} \small \it Department of Public Health Sciences, University of Rochester School of Medicine and Dentistry, Rochester, NY 14642, USA}
        \and
        \vspace{-2mm} EUN SUG PARK\thanks{To whom correspondence should be addressed: \texttt{e-park@tamu.edu}}\\
        {\vspace{-4mm} \small \it Texas A\&M Transportation Institute, Texas A\&M University System, College Station, TX 77843, USA}
        }
\date{January 2024}

\maketitle

\begin{abstract}
       Accounting for exposure measurement errors has been recognized as a crucial problem in environmental epidemiology for over two decades. 
    Bayesian hierarchical models offer a coherent probabilistic framework for evaluating associations between environmental exposures and health effects, which take into account exposure measurement errors introduced by uncertainty in the estimated exposure as well as spatial misalignment between the exposure and health outcome data.
    While two-stage Bayesian analyses are often regarded as a good alternative to fully Bayesian analyses when joint estimation is not feasible, there has been minimal research on how to properly propagate uncertainty from the first-stage exposure model to the second-stage health model, especially in the case of a large number of participant locations along with spatially correlated exposures. We propose a scalable two-stage Bayesian approach, called a sparse multivariate normal (sparse MVN) prior approach, based on the Vecchia approximation for assessing associations between exposure and health outcomes in environmental epidemiology. 
    We compare its performance with existing approaches through simulation. Our sparse MVN prior approach shows comparable performance with the fully Bayesian approach, which is a gold standard but is impossible to implement in some cases. We investigate the association between source-specific exposures and pollutant (nitrogen dioxide (NO$_2$))-specific exposures and birth outcomes for 2012 in Harris County, Texas, using several approaches, including the newly developed method.
\end{abstract}

\textit{Keywords}: Spatial exposure measurement error; Source-specific air pollution; Uncertainty propagation; Two-stage Bayesian approach; Vecchia approximation

\section{Introduction}
Modeling air pollution exposures and estimating health effects has been a long-standing challenge in air pollution epidemiology. Because ambient air pollutant measurements are only available at relatively few monitoring stations, different from the residences of study participants, exposure measurement error caused by spatial misalignment is introduced. In addition, when statistical models (such as spatiotemporal models, land use regression models, or source apportionment models) are used for exposure assessment, uncertainty in prediction also contributes to exposure measurement error. Previous studies show that failure to properly address exposure measurement error leads to biased health effect estimates as well as incorrect uncertainty estimates \citep{Carroll2006-jw,Dominici2000-gq,Zeger2000-dh,Gryparis2009-mm,Sheppard2012-oq,Park2014-iq}, which point to the need for the development of exposure and health effect models that account for these different sources of exposure measurement error. 

One research direction is to identify and/or decompose the measurement error into classical- and Berkson-type errors to obtain more accurate health effect estimates in a two-stage setting, for example, by using regression calibration~\citep{Van_Roosbroeck2008-bi}, parametric and nonparametric bootstrap methods~\citep{ Szpiro2011-ge,Keller2017-ts} or simulation extrapolation under spatially correlated measurement error \citep{Alexeeff2016-mp}. Bayesian hierarchical models, which do not require decomposition of the measurement error into the classical- and Berkson-type errors, have been employed as a coherent way to account for exposure measurement errors in health effects evaluation \citep{Molitor2006-xu,Nikolov2007-jh,Calder2008-on,Park2014-iq,Park2015-du, Park2018-iv}. 
When implementing a fully Bayesian approach is not feasible due to the complexity of the exposure model or a privacy issue in the use of health data prohibiting directly linking geocoded addresses with health data, an alternative strategy to account for the uncertainty of exposure estimates in the health effect estimates is to adopt a two-stage analysis: in the first-stage, an exposure model is fitted to obtain predictions at participant locations, and to then use those predictions in a second-stage health model.

Two-stage Bayesian approaches have received significant attention in recent years \citep{Gryparis2009-mm,Peng2010-qh,Chang2011-vb,Lee2017-tn}. Assuming correctly specified models, the resulting posterior distribution gives a valid estimate that is consistent with other advocated methods such as fully Bayesian approaches or regression calibration \citep{Gryparis2009-mm,Chang2011-vb}. 
We emphasize that two-stage Bayesian approaches are general and applicable not only to studies examining the health impact of pollutant-specific exposures (e.g. fine particulate matter (PM$_{2.5}$) or nitrogen dioxide (NO$_2$)) but also to studies where source-specific exposures (e.g., emissions from sources such as the transportation sector or refineries) are of interest. In the latter case, a source apportionment model such as Bayesian multivariate receptor models \citep{Park2014-iq,Park2018-iv,Park2018-sa,Hackstadt2014-hz,Park2020-za, Park2021-pi, Lee2023-cx, Baerenbold2022-dn}, can be used as an exposure model in the first stage.

Surprisingly, there has been little research in air pollution epidemiology on how to propagate uncertainty from an exposure model in the first stage to the second stage health model in two-stage Bayesian analyses, and the associated computational challenges have often been neglected. Recently, \citet{Comess2024-ot} provided a review of existing uncertainty propagation methods in two-stage analyses and proposed a kernel density estimation (KDE) approach to handle skewed exposure posterior predictive distributions, which can be useful in population-level time series analysis. However, the univariate version of KDE does not account for correlation between the exposures corresponding to different data points. While the multivariate KDE approach has the potential to account for correlation between exposures, it suffers from computational difficulties in selecting and evaluating the multivariate kernel even for a small number of data points (such as 250 as used in). Given that the number of participants in air pollution health effects studies is typically larger (e.g., several thousand to tens of thousands of individuals), accounting for spatial exposure measurement error in a two-stage Bayesian analysis is computationally challenging and often infeasible.

In this paper, we propose a sparse multivariate normal (sparse MVN) prior approach that can overcome the prohibitive computational burden for a large number of participant locations while efficiently propagating uncertainty due to spatial exposure measurement error in two-stage Bayesian analyses. The proposed sparse MVN prior approach gives a scalable solution to the uncertainty propagation problem with Vecchia approximation that yields sparse inverse covariance matrices. We compare its performance with several existing approaches by simulation. Through extensive simulation studies, we show that when the number of participant locations is large, the proposed approach can reduce computational costs by orders of magnitude and make the implementation feasible while sacrificing minimal dependency information and yielding accurate health effects and uncertainty estimates. We apply the methods to evaluate associations between air pollution and birth weight (modeled both as a continuous and dichotomous outcome) using birth certificate data from the Texas Department of State Health Services (TX DSHS) and air pollution data from the Texas Commission on Environmental Quality (TCEQ) for Harris County, Texas.

\section{Background}

\subsection{Bayesian methods for exposure and health modeling}
First, we introduce critical notation. Let $\calD$ be a spatial domain. Let $\{ s_1, \dots,s_{n_w}\}=\calS\subset \calD$ be a set of $n_w$ locations with exposure measurements $\bfW = [W(s_1),\dots,W(s_{n_w})]^\top$ that are subject to measurement error. 
Similarly, let $\{ s^*_1,\dots,s^*_{n_y} \}=\calS^* \subset \calD$ be a set of residential locations for $n_y$ participants with health outcomes $\bfY^* = [Y(s_1^*),\dots,Y(s^*_{n_y})]^\top$. 
Denote $\{X(s):s\in \calD\}$ be an unknown true exposure, assumed to be smoothly varying over the spatial domain $\calD$, and let $\bfX = [X(s_1),\dots,X(s_{n_w})]^\top$ and $\bfX^* = [X(s_1^*),\dots,X(s_{n_y}^*)]^\top$ be the true concentrations at exposure measurement and participant locations. Also, let $\bfZ^* = [\bfz(s_1^*),\dots,\bfz(s_{n_y}^*)]^\top$ be a $n_y\times p$ matrix of covariates in the health model that are not subject to measurement error. 

\begin{figure}
    \centering
    \includegraphics[width=0.5\textwidth]{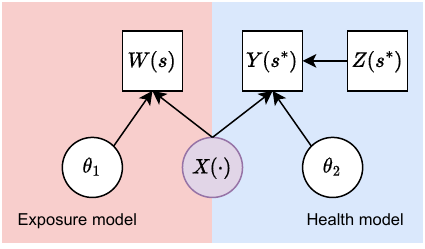}
    \caption{A diagram of exposure and health models, squares and circles indicate observable and latent variables, respectively. $\theta_1$ and $\theta_2$ indicate parameters of the first stage (exposure) and second stage (health) models. The exposure measurements $W(s)$ and health outcomes $Y(s^*)$ are spatially misaligned, which are jointly modeled with latent exposure surface $X(\cdot)$ defined on the whole domain. 
    }
    \label{fig:exposurehealth}
\end{figure}

The models describing exposures and the exposure-outcome association both involve true exposure surface $X(s)$ over the domain $\calD$; see Figure~\ref{fig:exposurehealth}. 
The spatial misalignment of exposure measurement locations $\calS$ and participant locations $\calS^*$ underscores the importance of uncertainty quantification because there is an increased uncertainty at unmonitored locations caused by predicting true exposure at participant locations $\bfX^*$ from the outdoor air pollutant measurements $\bfW$ obtained from monitoring locations. 
To this end, we follow a Bayesian framework to jointly model $(\bfX,\bfX^*)$ by introducing a Gaussian process (GP) prior on the true exposure surface $\{X(s):s\in \calD\}$:
\begin{align}
     \text{Exposure measurement model:}&\quad W(s_h) = X(s_h) + \delta(s_h), \quad \delta(s_h)\iidsim \mathrm{N}(0,\sigma_W^2), \label{eq:firststage}\\
     \text{Health model:}&\quad \bbE[Y(s_i^*)\given X(s_i^*)] =  g^{-1}(\beta_0 + \beta_xX(s_i^*) + \bfz(s_i^*)^\top \bm\beta_z ), \label{eq:secondstage}\\
      \text{Prior on $X$:}&\quad X(\cdot)\sim \mathrm{GP}(\mu_X(\cdot), k_X(\cdot,\cdot))  \label{eq:Xprior}, 
\end{align}
for $h=1,\dots,n_w$ and $i=1,\dots,n_y$, where $\beta_x$ is the health effect parameter that is of the primary interest,  $\beta_0$ and $\bm\beta_z$ are an intercept and a vector of covariate coefficients, respectively, and $\mu_X(\cdot)$ and $k_X(\cdot,\cdot)$ are the mean and covariance kernel of GP.  
Here $g$ is the link function that accommodates different types of health outcomes under the generalized linear model framework. For example, when $Y$ is continuous real-valued data, letting $g(\mu) = \mu$ becomes \eqref{eq:secondstage} a linear regression model. When $Y$ is binary data, letting $g(\mu) = \mathrm{logit}(\mu) = \ln(\mu/(1-\mu))$ becomes \eqref{eq:secondstage} a logistic regression model. 
Throughout the paper, we will focus on two canonical health model examples: (i) the normal linear regression model for a continuous health outcome, and (ii) the logistic regression model for a binary health outcome.

As discussed earlier, a two-stage Bayesian approach is often employed as an alternative to a fully Bayesian approach in practice. That is, a researcher first fits the exposure model \eqref{eq:firststage}, predicts $\bfX^*$ based on $\bfW$, then subsequently fits the health model \eqref{eq:secondstage} using the prediction from the first stage.
It is important to propagate uncertainty associated with $\bfX^*$ to the second stage to properly quantify the uncertainty associated with the health effect estimate $\beta_x$, which will be described next.

\subsection{Comparison of two-stage Bayesian approaches}
From the first stage Bayesian exposure model, we obtain the posterior predictive distribution of $n_y$-dimensional $\bfX^*$, typically represented as a set of $N$ samples $\{\bfX^{*(\ell)}\}_{\ell=1}^N$ generated from Markov chain Monte Carlo (MCMC) methods. 
We outline three approaches that are often used for a two-stage Bayesian analysis in environmental epidemiology.

\begin{enumerate}
    \item \textit{Plug-in}. The plug-in approach summarizes posterior predictive samples with a point estimate $\hat\bfX^*$, such as sample mean $\hat\bfX^* = \bfm = (m_1,\dots,m_{n_y}) = N^{-1}\sum_{\ell=1}^N\bfX^{*(\ell)}$, and then those values are plugged in the second stage model \eqref{eq:secondstage}; see \citet{Warren2022-cg} for an example. 
    \item \textit{Independent normal prior}. The independent normal prior approach summarizes posterior predictive samples with independent normals. That is, the second stage prior density $p(\bfX^*)$ is simply a product of $p(X(s_i^*))$, each with a normal distribution having mean $m_{i}$ and variance $(N-1)^{-1}\sum_{\ell=1}^N (X(s_i^*)^{(\ell)}-m_{i})^2$ for $i=1,\dots,n_y$. This approach takes account of marginal variances but drops spatial dependency information for simplification; see \citet{Huang2018-yd,Cameletti2019-wc}.
    \item \textit{Multivariate normal prior (MVN prior)}. The MVN prior approach summarizes posterior predictive samples with a multivariate normal distribution. The second stage prior is a $n_y$-dimensional multivariate normal $\bfX^*\sim \mathrm{N}_{n_y}(\bfm, \bfS)$ with sample mean and covariance matrix,
    \begin{equation}
    \label{eq:mvnmeancov}
    \bfm = \frac{1}{N}\sum_{\ell=1}^N \bfX^{*(\ell)}, \qquad \bfS = \frac{1}{N-1}\sum_{\ell=1}^N (\bfX^{*(\ell)}-\bfm)(\bfX^{*(\ell)}-\bfm)^\top.
    \end{equation}
    The MVN prior approach fully takes second-order dependency information into account (see \citet{Warren2012-tc,Lee2017-tn}, for examples).
\end{enumerate}

It is well known that each approach yields different inferential results for the health effect analysis in the second stage \citep{Gryparis2009-mm}. For example, the plug-in approach ignores the uncertainty from the first stage analysis and often results in undercoverage of the true health effect parameters. 
The MVN prior approach can be viewed as an MVN approximation of the fully Bayesian model based on posterior predictive samples $\bfX^{*(1)},\dots,\bfX^{*(N)}$. 
Although the independent normal prior approach is often adopted as a simplified version of the MVN prior approach to reduce computational burden, it cannot account for spatial correlation across exposures and its effect on estimation of the health effect parameter is not yet well understood. 
There are also other approaches, such as a \textit{discrete uniform prior} \citep{Peng2010-qh,Chang2011-vb} and \textit{exposure simulation} \citep{Blangiardo2016-cq}, but the former makes an unrealistic assumption that true exposure is contained in the samples \citep{Comess2024-ot}, and the latter is known to produce a biased estimate \citep{Gryparis2009-mm}. 

\subsection{Existing computational challenges of the MVN approach}

Among the uncertainty propagation approaches outlined in the previous section, the MVN prior approach is known to have the least bias on health effect estimates and well approximates the fully Bayesian approach \citep{Gryparis2009-mm}.  
Also, representing second stage prior $p(\bfX^*)$ as an $n_y$-dimensional MVN allows fitting the second stage model with a Gibbs sampler under the Bayesian normal linear health model \eqref{eq:secondstage} as well as probit, logistic, and negative binomial regression models using data augmentation strategies \citep{Albert1993-js,Polson2013-gb}. That is, under the MVN prior $\bfX^* \sim \mathrm{N}_{n_y}(\bfm, \bfS)$, we draw random samples of $\bfX^*$ from its full conditional distribution, which is also an $n_y$-dimensional MVN,
\begin{equation}
\label{eq:xfullcond}
    \bfX^* \given - \sim \mathrm{N}_{n_y}\left((\bfS^{-1} + \bfD)^{-1}\bfb, (\bfS^{-1} + \bfD)^{-1}\right),
\end{equation}
with some vector $\bfb\in \bbR^{n_y}$ and $n_y\times n_y$ covariance matrix $(\bfS^{-1} + \bfD)^{-1}$ containing a diagonal matrix $\bfD$. For instance, under a normal linear model with error variance $\sigma_Y^2$, we have $\bfD = (\beta_x^2/\sigma_Y^2)\bfI_{n_y}$; see Appendix \ref{appendix:postinferencedetail} for a description of a logistic regression model as well.

However, the MVN prior approach faces significant computational challenges when the number of participant locations $n_y$ becomes large. 
Recall that the second stage posterior inference algorithm with the MVN prior approach involves repeatedly drawing a random sample of $\bfX^*$ from an 
  $n_y$-dimensional MVN \eqref{eq:xfullcond}, where $n_y$ may be more than several thousand or tens of thousands for many real scenarios. In such cases, second-stage inference becomes practically infeasible because sampling from MVN takes the cubic complexity of its dimension \citep[][\S 4.1.2 and \S 4.2.3]{Golub2013-pe}. An existing MVN prior approach sequentially updates each $n_y$ coordinate of $\bfX^*$ one at a time using univariate normal full conditionals \citep{Lee2017-tn}, but it puts convergence of MCMC algorithm and computational efficiency in serious jeopardy because each element of $\bfX^*$ corresponds to a spatial location and thus the elements of $\bfX^*$ are typically highly correlated.

When the inverse covariance (precision) matrix $\bfS^{-1} + \bfD$ in \eqref{eq:xfullcond} is sparse so that it has many zero elements, there exist very efficient MVN sampling algorithms by exploiting the sparse structure of the precision matrix based on sparse Cholesky factorization \citep{Rue2001-nc,Furrer2010-ky}.
Then, the computational complexity of MVN sampling depends primarily on the number of nonzero elements of the precision matrix instead of its dimension, which provides a significant computational benefit compared to standard algorithms that do not exploit sparsity. 
This motivates us to replace $\bfS^{-1}$ with some suitable sparse matrix $\bfQ$, and we propose to use $\bfX^*\sim \mathrm{N}_{n_y}(\bfm, \bfQ^{-1})$ as a prior instead of the original MVN prior $\bfX^*\sim \mathrm{N}_{n_y}(\bfm, \bfS)$, which has a dense precision matrix referred as ``dense MVN''. The question is how to choose the sparse matrix $\bfQ$ appropriately so that the sparse MVN prior is close to the original dense MVN prior that contains spatial dependency information in order to minimize its impact on estimation in the second-stage health model. The independent normal prior approach can also be considered as a sparse MVN prior approach by choosing $\bfQ$ be a diagonal matrix with elements $\var(X(s_i^*))^{-1}$, $i=1,\dots,n_y$, but it discards all spatial dependency information.

\section{Uncertainty propagation with sparse MVN prior}

\subsection{Overview of Vecchia approximation}

\begin{figure}
     \centering
     \begin{subfigure}[b]{0.32\textwidth}
         \centering
         \includegraphics[width=\textwidth]{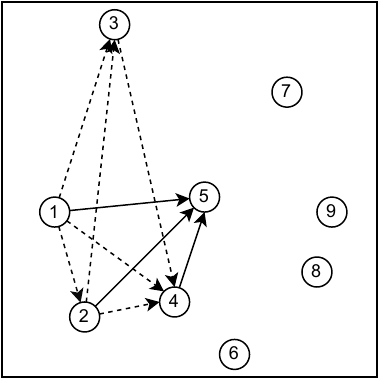}
     \end{subfigure}
     \hfill
     \begin{subfigure}[b]{0.32\textwidth}
         \centering
         \includegraphics[width=\textwidth]{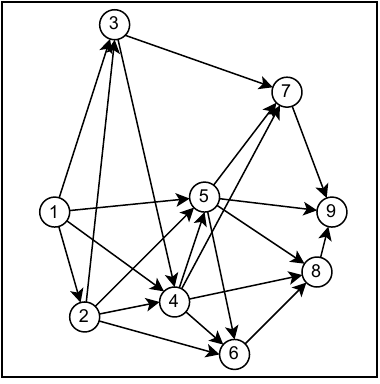}
     \end{subfigure}
     \hfill
     \begin{subfigure}[b]{0.32\textwidth}
         \centering
         \includegraphics[width=\textwidth]{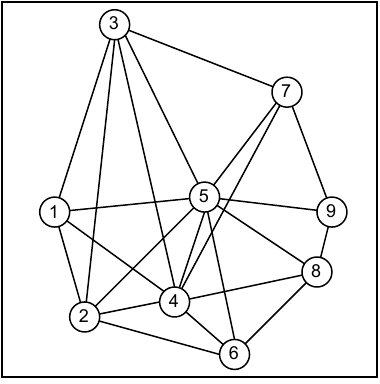}
     \end{subfigure}
        \caption{Vecchia approximation with left-to-right ordering shown in numbers and 3 nearest neighborhood conditioning sets. (Left panel) Conditioning set relation of $N(s_5) = \{s_1,s_2,s_4\}\subset \{s_1,s_2,s_3,s_4\}$ shown in black arrows, and previous conditioning sets shown in dashed arrows. (Center panel) Resulting directed acyclic graph (DAG) that represents the sparsity pattern of the reverse Cholesky factor $\bfU$. (Right panel) The Gaussian Markov random field induced by Vecchia approximation, the moral graph of a DAG that represents conditional independence structure as well as sparsity pattern of the precision matrix $\bfQ$. 
        Note that an edge $(s_3-s_5)$ has been added since both have a common child node $s_7$.
        }
        \label{fig:vecchia}
\end{figure}

The Vecchia approximation \citep{Vecchia1988-gf} is a collection of approximation methods for the MVN distribution. Let $p(\bfX^*) = p(X^*(s_1), \dots, X^*(s_n))$ be a $n$-dimensional MVN density with mean $\bfm$ and covariance $\bfS$ associated with locations $s_1,\dots,s_{n}$.  Based on the product of conditional distribution representation of a joint distribution, \citet{Vecchia1988-gf} proposed to approximate the MVN distribution by truncating the conditioning variables,
\begin{align}
  p(\bfX^*) = p(X^*(s_1))\prod_{i=2}^n p(X^*(s_i)\given X^*(s_1),\dots,X^*(s_{i-1}) ) \approx p(X^*(s_1))\prod_{i=2}^n p(X^*(s_i)\given \bfX^*_{N(s_i)})
\end{align}
where $N(s_i)\subset \{s_1,\dots,s_{i-1}\}, i=2,\dots,n$ are called conditioning sets and $\bfX^*_{N(s_i)}$ is the vector formed by stacking the realizations of $X^*(s)$ over $N(s_i)$. By the properties of MVN, each conditional distribution $p(X^*(s_i)\given \bfX^*_{N(s_i)})$ is normal and the resulting approximated density is also MVN with the same mean but with different covariance. 
The ordering of data indices and the choice of conditioning sets determine the Vecchia approximation scheme, and the approximation is exact when $N(s_i)= \{s_1,\dots,s_{i-1}\}$ for all $i=2,\dots,n$. 
The most popular way to choose conditioning sets is based on $k$ nearest neighbors with small~$k$, the scheme that Vecchia originally proposed.
See Figure~\ref{fig:vecchia} (left panel) for an illustration with a left-to-right ordering and $3$ nearest neighbor conditioning sets, where conditioning set relations are represented with arrows (directed edges), denoted as $(s_i\to s_j)$ if $s_i\in N(s_j)$. Then, the center panel of Figure~\ref{fig:vecchia} shows all the conditioning set relations, forming a directed acyclic graph (DAG).

The most important feature of Vecchia approximation is that the resulting precision matrix $\bfQ$ admits sparse Cholesky decomposition $\bfQ = \bfU \bfU^\top$.
Here $\bfU$ is an upper triangular sparse matrix, where $(s_i\to s_j)$ is equivalent to $U_{ij} \neq 0$ for $i<j$, so that the DAG in the center panel of Figure~\ref{fig:vecchia} illustrates a sparsity pattern of $\bfU$. 
The number of off-diagonal nonzero elements for each column of $\bfU$ corresponds to the size of the conditioning set $N(s_i)$; see \citet{Finley2019-zw, Katzfuss2021-nt} for details.
Thus, by choosing a small number of neighborhoods $k$ for the conditioning set, the corresponding Cholesky factor $\bfU$ is highly sparse, which leads to a sparse precision matrix $\bfQ$, motivating many recent scalable Gaussian process methodologies \citep{Datta2016-dx,Peruzzi2022-yf,Quiroz2023-lg}.

In an epidemiological application, the interpretation of the conditional independence structure induced by the Vecchia approximation is worth discussing. 
 The MVN with precision matrix $\bfQ$ can be considered as a Gaussian Markov random field (GMRF), where $Q_{ij} = 0$ is equivalent to the $i$th and $j$th variables being conditionally independent given all other variables; see \cite{Rue2005-er}
 for a comprehensive overview. 
 As shown in the right panel of Figure~\ref{fig:vecchia}, the sparsity pattern of $\bfQ$ can be visualized with an undirected graph where an undirected edge $(s_i-s_j)$ corresponds to $Q_{ij} \neq 0$ for $i\neq j$. In fact, the undirected graph associated with GMRF is the \textit{moral graph} of DAG \citep[\S 4.5]{Koller2009-qz}, both representing the same conditional independence structure.

\subsection{Sparse MVN approach with Vecchia approximation}

To address the computational challenge in Bayesian two-stage exposure-health analysis with a large number of participants, we propose the \textit{sparse MVN prior} approach based on Vecchia approximation as a scalable alternative to the \textit{dense MVN prior} approach. That is, given the ordering of data and the choice of conditioning sets, we propose to use prior $\bfX^*\sim \mathrm{N}_{n_y}(\bfm, \bfQ^{-1})$ with a sparse precision matrix $\bfQ$ obtained from a Vecchia approximation that well approximates a dense MVN prior $\bfX^*\sim \mathrm{N}_{n_y}(\bfm, \bfS)$. While there are many possible ways to choose ordering and conditioning sets, we follow the configuration of nearest neighbor GP \citep{Datta2016-dx} that suggests coordinate-based ordering and $k$-nearest neighbor conditioning sets. 
\citet{Datta2016-dx} reported the results are ``extremely robust to the ordering'' with the nearest neighbor conditioning sets.

Based on the sparse precision matrix $\bfQ$ obtained from a Vecchia approximation, the sparse MVN prior  $\bfX^*\sim \mathrm{N}_{n_y}(\bfm, \bfQ^{-1})$ leads to the full conditional of $\bfX^*$ in \eqref{eq:xfullcond} having a sparse precision matrix $\bfQ + \bfD$, allowing to use an efficient MVN sampling algorithm even when dimensionality $n_y$ is very high. One may be concerned about the computational cost of finding $\bfQ$ itself, but when the size of the conditioning set is bounded by small $k$ (such as in the nearest neighbor conditioning scheme), finding $\bfQ$ only takes linear complexity in its dimension $n_y$ \citep{Finley2019-zw}. 

We conduct a simulation study to analyze how the quality of the Vecchia approximation differs by the choice of conditioning sets and investigate how much computational gain comes with an approximation. We generate $n = 1000, 2000,\dots, 5000$ points uniformly at random on a spatial domain $\calD = [0,2]\times [0,2]$, and consider an $n$-dimensional mean zero MVN distribution $p_n$ with exponential covariance function $k(s_i,s_j) = \exp(-\|s_i - s_j\|_2)$, which yields
a dense precision matrix. 
We apply the Vecchia approximation to $p_n$ using coordinate-based ordering and $k$-nearest neighborhood conditioning, and we denote the approximated distribution as $\tilde{p}_n^{(k)}$. To measure the quality of the approximation, we compare the Kullback-Leibler (KL) divergence $D_\mathrm{KL}\left(p_n \,||\, \tilde{p}_n^{(k)}\right)$ as $k$ varies from $k=0,3,5,$ (here $k=0$ corresponds to an empty conditioning set that leads to an independent normal approximation of $p_n$). To assess the computational benefit, we compare the wall-clock time to draw one sample from a multivariate normal distribution. We used R package \texttt{mvnfast} \citep{Fasiolo2014-dx} for drawing samples from dense MVN and R package \texttt{spam} \citep{Furrer2010-ky} for sparse MVN. All computations were performed on Intel E5-2697 v2 CPU with 128GB of memory.

\begin{table}
    \centering
    \small
   \caption{Average KL divergence as an approximation quality assessment and MVN sampling time per sample based on 100 replicates (both lower the better). Sparse MVN ($k$nn) corresponds to Vecchia approximation with coordinate-based ordering and $k$ nearest neighbor conditioning.}
   \begin{tabular}{c c ccccc}
    \toprule
      & Distribution &  $n=1000$ & $n=2000$ & $n=3000$ & $n=4000$ & $n=5000$\\
     \midrule
    \multirow{3}{*}{$D_{\mathrm{KL}}\left(p_n \,||\, \tilde{p}_n^{(k)}\right)$} & Independent normal & 1452.7 & 3260.2 & 5203.0 & 7228.1& 9319.7 \\
 & Sparse MVN (3nn) &47.9 & 104.0 & 161.5& 219.9 & 279.0 \\ 
 & Sparse MVN (5nn) &23.4 & 51.9 & 81.7& 111.7 & 142.2  \\
     \midrule
    \multirow{4}{*}{\makecell{Sampling time \\ (sec)} } &  Independent normal & 0.0006 & 0.0009 & 0.0011 & 0.0014 & 0.0017\\
 & Sparse MVN (3nn) & 0.0012 & 0.0022 & 0.0032 & 0.0042& 0.0055 \\
 & Sparse MVN (5nn) & 0.0017 & 0.0036 & 0.0052 & 0.0075 & 0.0102 \\
 & Dense MVN ($p_n$) & 0.0238 & 0.1625 & 0.4788 & 1.0666 & 2.0364\\ 
     \bottomrule
  \end{tabular}
    \label{tab:vecchia}
\end{table}

The results based on 100 replicates are summarized in Table~\ref{tab:vecchia}. The independent normal $p_n^{(0)}$ is a poor approximation of $p_n$ in terms of KL divergence, but the Vecchia approximation provides a significant improvement even with a small number of conditioning sets. When it comes to the increasing rate of MVN sampling time as $n$ increases from 1000 to 5000, the independent and sparse MVN sampling times only increased about 3-, 5-, or 6-fold, but the dense MVN sampling time increased more than 80-fold due to its cubic complexity. 
The wall-clock time taken for Vecchia approximation (i.e., the time for finding the sparse precision matrix $\bfQ$ itself), which only needs to be performed once, took less than 10 seconds in all cases.
The results suggest that when the number of participant locations is more than tens of thousands, conducting a two-stage Bayesian inference based on the dense MVN approach is not feasible in practice.
As an alternative, the \textit{sparse MVN} approach provides a reasonable solution that balances approximation quality and computational benefits.
We emphasize that the sparse MVN prior approach based on the Vecchia approximation has not yet been utilized in the context of two-stage exposure-health modeling, which enables us to perform uncertainty propagation with spatial correlations taken into account even when there is a large number (tens of thousands) of participants with health data.

\section{Simulation Studies}

In this section, in the context of two-stage Bayesian exposure-health analysis, we conduct a simulation study to analyze how the proposed sparse MVN prior approach leads to different health effect estimation results compared to other approaches. 
\subsection{Simulation settings}

We describe the first-stage exposure model that we used in the simulation studies. We consider the discrete process convolution (DPC) model \citep{Higdon1998-pr,Higdon2002-gi}, one of the simple low-rank Gaussian process methods by modeling the true exposure surface $X(\cdot)$ as
\begin{equation}
    X(s) = \mu + \sum_{l=1}^L K(s-u_l)G(u_l), \quad s\in\calD,
    \label{eq:dpc}
\end{equation}
where $\mu \in\bbR$ represents the overall mean, $K(\cdot-u_l)$ is a smoothing kernel centered at fixed grid locations $u_l\in\calD$ for $l = 1,\dots,L$, and $\bfG = (G(u_1),\dots,G(u_L))^\top$ is a latent discrete process with prior $\bfG\sim \mathrm{N}_L(\bm{0}, \sigma_G^2\bfI_L)$. 
Specifically, we use a bivariate normal kernel $K(s-u_l) = (2\pi\sigma_k^2)^{-1}\exp\left(-\|s-u_l\|_2^2/(2\sigma_k^2)\right)$ with standard deviation $\sigma_k$ equal to the distance between adjacent grid locations. Combined with the measurement error model \eqref{eq:firststage}, the first stage model can be written as $\bfW = \mu\bm{1}_{n_w} + \bfK\bfG + \bm\delta$ with error $\bm\delta \sim \mathrm{N}_{n_w}(\bm{0}, \sigma_W^2\bfI)$, where $\bfK\in\bbR^{n_w\times L}$ is a convolution kernel matrix with $(h,l)$-th element $K(s_h-u_l)$.
Next, for the second stage health model, we consider continuous and binary health outcomes. For continuous $Y(s^*_i)$, we fit the Bayesian normal linear regression model with intercept, exposure $X(s_i^*)$, and a covariate $Z(s_i^*)$ as predictors. For binary health outcomes, we fit the Bayesian logistic regression model with the same predictors. 
While there are other options for first-stage exposure modeling \citep{Banerjee2008-iw,Finley2009-bc,Cressie2008-rs,Datta2016-dx,Katzfuss2017-lo}, we adopt the DPC model in the first stage because it is not only simple, scalable, and easily amenable to accommodating nonstationarity of the exposure surface through selection of a spatially-varying kernel \citep{Paciorek2006-tl}, but also allows for a seamless combination of the first and second stage models (when there is not an issue of temporal misalignment) to carry out fully Bayesian inference. 

\begin{figure}
    \centering
    \includegraphics[width=0.9\textwidth]{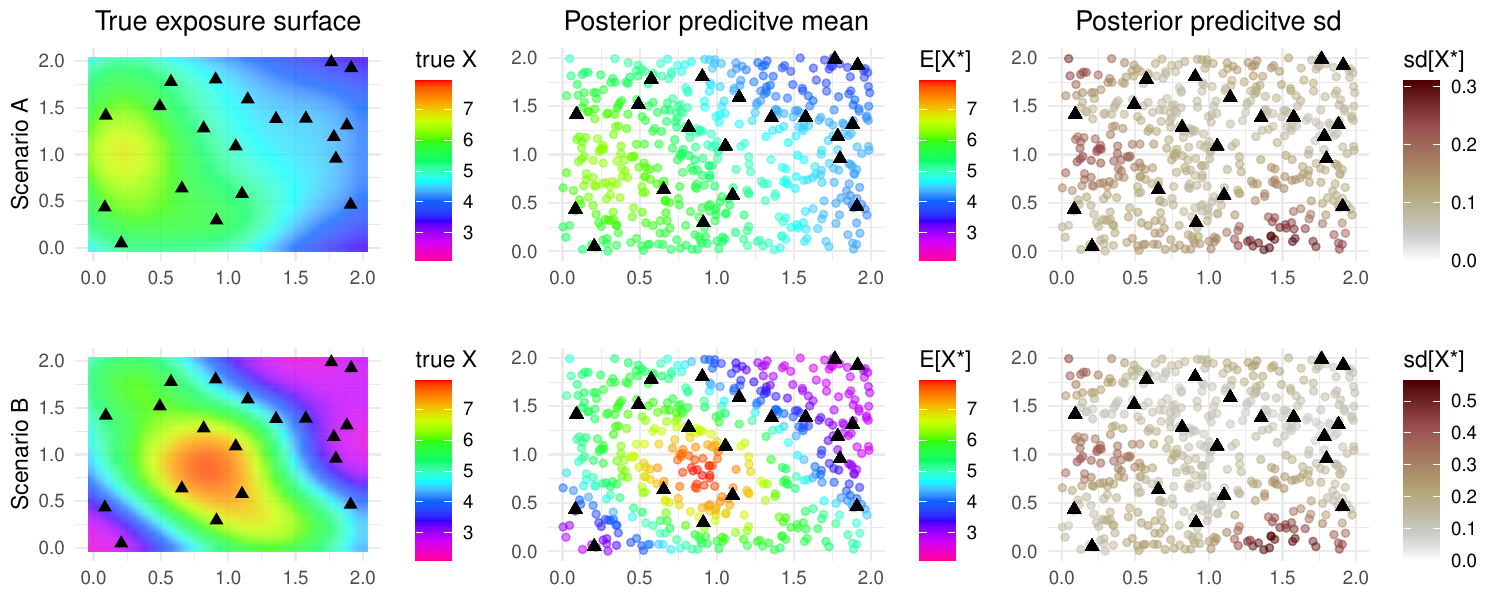}
    \caption{(Left panel) Two different scenarios of the true exposure surface. Triangles indicate $n_w= 20$ exposure measurement locations. (Center and Right panels) Posterior predictive mean and standard deviation (sd) at participant locations from the first stage fit. Circles indicate $n_y = 500$ participant locations.}
    \label{fig:scenarioAB}
\end{figure}

The simulated data are generated as follows. Similar to \cite{Gryparis2009-mm}, we consider two different scenarios of the true exposure surface $X(\cdot)$, Scenario A with a smoother exposure surface and Scenario B with a more heterogeneous surface. 
On a spatial domain $\calD = [0,2]^2$, we generate $X$ from the DPC model \eqref{eq:dpc} with different variability as shown in Figure~\ref{fig:scenarioAB};  Appendix~\ref{appendix:detailsettings} provides further details. 
We assign $n_w = 20$ exposure measurement locations $\{s_1,\dots,s_{n_w}\}$ which are uniformly distributed over $\calD$, and generate exposure measurements $W(s_1),\dots,W(s_{n_w})$ with measurement error  $\bm\delta \sim \mathrm{N}_{n_w}(\bm{0}, 0.1^2\bfI)$. 
The participant locations $\{s_1^*,\dots,s_{n_y}^*\}$ are also uniformly distributed over the domain with sizes $n_y \in \{1000,2000,5000\}$. 
The covariates $Z(s_i^*)$, $i=1,\dots,n_y$, which are not subject to measurement error, are generated from the uniform distribution between 0 and 1 and fixed thereafter. Finally, we consider a normal linear model for a continuous health outcome:
\begin{equation}
\label{eq:contioutcome}
    Y(s_i^*) = \beta_0 +  \beta_x X(s_i^*) + \beta_z Z(s_i^*) + \epsilon_i, \quad \epsilon_i\iidsim \mathrm{N}(0, \sigma_Y^2) , \quad i=1,\dots,n_y,
\end{equation}
which corresponds to \eqref{eq:secondstage} with identity link $g(\mu) = \mu$. We set $\beta_0 = 0$, $\beta_x = 1$, $\beta_z = 2$, and $\sigma_Y^{2} = 0.64$ in the data generating step. 
Similarly, we consider a logistic regression model for a binary outcome: 
\begin{equation}
\label{eq:binaryoutcome}
    \mathrm{logit}(\bbP(Y(s_i^*)=1 \given X(s_i^*))) = \beta_0 + \beta_x X(s_i^*) +\beta_zZ(s_i^*), \quad i=1,\dots, n_y, \text{ independently,}
\end{equation}
which corresponds to \eqref{eq:secondstage} with logit link $g(\mu) = \ln(\mu/(1-\mu))$. We set $\beta_0 = -7$, $\beta_x = 1$, and $\beta_z = 2$ in data generation. 
Based on this setting, we generated a total of 400 simulated datasets.

We compare several different methods in the simulation study. First is the \textit{plug-in} approach, where the posterior predictive mean $\bfm$, visualized in the center panel of Figure~\ref{fig:scenarioAB}, is plugged into $\bfX^*$ in the second stage model, without incorporating any uncertainty information. Second is the \textit{independent normal prior} approach, where the prior $\bfX^*\sim \mathrm{N}_{n_y}(\bfm,\diag(\bfS))$ is adopted in the second stage by taking only diagonal elements of the sample covariance matrix $\bfS$, visualized in the right panel of Figure~\ref{fig:scenarioAB}. Next is the proposed \textit{sparse MVN prior} approach with two different conditioning sets, where the prior $\bfX^*\sim \mathrm{N}_{n_y}(\bfm,\bfQ^{-1})$ based on the Vecchia approximation with left-to-right ordering and $k=3$ and $k=5$ nearest neighbor conditioning sets are considered. The \textit{dense MVN prior} approach corresponds to the case when the dense MVN prior $\bfX^*\sim \mathrm{N}_{n_y}(\bfm,\bfS)$ is used, and the \textit{fully Bayesian} approach corresponds to the model that jointly fits the first and second stages under the DPC framework. 
Finally, we have added \textit{True exposure} when true exposure $\bfX^*$ is plugged in the second stage health model for comparison. Further details on the choice of prior distributions, MCMC algorithms, and other settings can be found in Appendices \ref{appendix:detailsettings} and  \ref{appendix:postinferencedetail}.

\subsection{Simulation results}

The simulation results are summarized in terms of the health effect estimate $\beta_x$. To examine the differences in estimated health effects across various approaches, we consider Bias, RMSE, average length of the credible interval, empirical coverage probability, and wall-clock time to fit the second-stage health model. Specifically, we consider posterior mean estimator $\hat\beta_x$ and calculate $\mathrm{Bias} = \bbE[\hat\beta_x] -\beta_x^{\mathrm{true}}$ and $\mathrm{RMSE} = (\bbE[(\hat\beta_x- \beta_x^{\mathrm{true}})^2])^{1/2}$ where expectation is taken over 400 simulated datasets. The second stage fitting time, omitted for the fully Bayesian approach, is based on the wall-clock time to run the MCMC algorithm with 20,000 iterations. 

\begin{table}[t]
  \small
  \centering
  \caption{Simulation results of a Bayesian linear regression model with a continuous outcome based on 400 replicates. Bias and RMSE are calculated with the posterior mean estimator $\hat\beta_x$. The $\bbE[\ell_{0.95}]$ and Coverage indicate the average length and empirical coverage of the 95\% credible interval of $\beta_x$. Time corresponds to the wall-clock time to fit the second stage model.}
  \label{table:conti1000}
  \vspace{3mm}
  \begin{tabular}{cc c c c c c c c}
    \toprule
    Continuous outcome & Method & Bias & RMSE & $\bbE[\ell_{0.95}]$ & Coverage(\%) & Time (s)\\
     \midrule
    \multirow{7}{*}{\makecell{Scenario A\\$n_y = 1000$}} & True exposure&  -0.002 & 0.032 & 0.126 & 95.0\% & 7.1\\
 & Plug-in & 0.040 & 0.101 & 0.140 & 60.2\% & 7.8\\
 & Independent normal & 0.033 & 0.093 & 0.137 & 63.0\% & 49.7\\ 
 & Sparse MVN (3nn) & -0.012 & 0.066 & 0.243 & 92.8\% & 74.7\\ 
 & Sparse MVN (5nn) & -0.005 & 0.065 & 0.282 & 96.0\% & 105.4\\ 
 & Dense MVN & 0.017 & 0.058 & 0.244 & 95.8\% & 1475.0\\ 
 & Fully Bayesian  & 0.012 & 0.055 & 0.226 & 96.5\% & -\\ 
     \midrule
    \multirow{7}{*}{\makecell{Scenario B\\$n_y = 1000$}} & True exposure &  0.000 & 0.018 & 0.069 & 95.0\% & 7.0\\
 & Plug-in & 0.017 & 0.053 & 0.076 & 53.5\% & 7.4\\
 & Independent normal & 0.014 & 0.045 & 0.074 & 60.5\% & 49.4\\ 
 & Sparse MVN (3nn) & 0.001 & 0.034 & 0.133 & 93.2\% & 77.7\\ 
 & Sparse MVN (5nn) & -0.004 & 0.036 & 0.147 & 96.0\% & 104.9\\ 
 & Dense MVN & 0.006 & 0.029 & 0.133 & 97.8\% & 1484.3\\ 
 & Fully Bayesian  &0.007 & 0.029 & 0.120 & 97.0\% & -\\ 
     \bottomrule
  \end{tabular}
\end{table}

\begin{table}[t]
  \small
  \centering
  \caption{Simulation results of a Bayesian logistic regression model with a binary outcome with $n_y = 5000$ based on 400 replicates. Bias and RMSE are calculated with the posterior mean estimator $\hat\beta_x$. The $\bbE[\ell_{0.95}]$ and Coverage indicate the average length and empirical coverage of the 95\% credible interval of $\beta_x$. Time corresponds to the wall-clock time to fit the second stage model.}
  \label{table:binary5000}
  \vspace{3mm}
  \begin{tabular}{cc c c c c c c c}
    \toprule
    Binary outcome & Method & Bias & RMSE & $\bbE[\ell_{0.95}]$ & Coverage & Time (s)\\
     \midrule
    \multirow{8}{*}{\makecell{Scenario A\\$n_y = 5000$}} & True exposure &  0.000 & 0.044 & 0.176 & 96.0\% & 107.9\\
 & Plug-in &0.033 & 0.103 & 0.186 & 69.8\% & 107.7 \\
 & Independent normal & 0.042 & 0.106 & 0.192 & 68.8\% & 158.3\\ 
 & Sparse MVN (3nn) &0.015 & 0.076 & 0.231 & 87.5\% & 258.2\\ 
 & Sparse MVN (5nn) & 0.002 & 0.071 & 0.265 & 94.8\% & 389.0\\ 
 & Dense MVN & 0.016 & 0.066 & 0.284 & 97.8\% & 48220.9\\ 
 & Fully Bayesian & 0.014 & 0.066 & 0.273 & 96.8\% & - \\ 
     \midrule
    \multirow{8}{*}{\makecell{Scenario B\\$n_y = 5000$}} & True exposure &  0.003 & 0.033 & 0.128 & 94.5\% & 108.5\\
 & Plug-in & 0.007 & 0.069 & 0.129 & 68.2\% & 108.4\\
 & Independent normal & 0.028 & 0.071 & 0.136 & 68.2\% & 157.2\\ 
 & Sparse MVN (3nn) & 0.015 & 0.053 & 0.160 & 90.0\% & 258.2\\ 
 & Sparse MVN (5nn) &0.011 & 0.051 & 0.178 & 93.0\% & 389.1\\ 
 & Dense MVN & 0.010 & 0.045 & 0.186 & 97.0\% & 48418.7\\ 
 & Fully Bayesian  & 0.010 & 0.045 & 0.178 & 96.8\% & -\\ 
     \bottomrule
  \end{tabular}
\end{table}

The simulation study results for scenarios A and B are summarized in Table~\ref{table:conti1000} for the regression model with a continuous outcome ($n_y =1000$), and Table~\ref{table:binary5000} for the regression model with a binary outcome ($n_y = 5000$). 
Additional results are reported in Table~\ref{table:conti20005000} and Table~\ref{table:binary10002000} in Appendix~\ref{appendix:moretable}. 
The results for the existing approaches are generally consistent with the previous study of \citet{Gryparis2009-mm}, and here we summarize the main findings. 
First, the plug-in and independent normal prior approaches both show high RMSE and poor coverage probability. 
While the poor performance and incorrect coverage probability of the plug-in approach are expected phenomena, the independent normal prior approach also suffers from high RMSE and low coverage probability, suggesting that propagating spatial correlation of exposure estimates through the prior plays an important role. 
The dense MVN prior approach and fully Bayesian approach generally lead to the lowest RMSE and reasonable empirical coverage probability. However, as shown in Table~\ref{table:binary5000}, the dense MVN prior approach takes a significantly longer time to fit the second stage model as compared to the sparse MVN prior approach, which is especially noticeable when $n_y = 5000$ (13 hours versus $<7$ minutes, respectively). This result suggests that the dense MVN prior approach is not feasible when $n_y$ is in the order of tens of thousands of observations.

The proposed sparse MVN prior approach provides a nice compromise between the independent normal prior and dense MVN prior approaches, where RMSE and coverage probability become better as the size of conditioning sets increases. 
By exploiting the sparsity structure obtained from the Vecchia approximation, the second stage Bayesian inference using the sparse MVN prior approach is much more scalable compared to the dense MVN prior approach, especially when the number of participant locations is large. In our simulation study, the 5-nearest neighborhood conditioning set provides a satisfactory result, but a larger conditioning set may be adopted if desired. 
Our findings are similar in both Scenarios A and B, but there are some small differences. Compared to Scenario B where the exposure surface is rough, the smooth exposure surface in Scenario A leads to high collinearity between the intercept and exposure terms, which generally leads to a larger bias and RMSE, and wider credible intervals. This may seem contradictory to the simulation results of \citet{Gryparis2009-mm}. Note, however, that the simulation setting of \citet{Gryparis2009-mm}, which generated exposure measurements ($W$) assuming exposure measurements are the same as the true exposures ($X$) and that the true exposures include local heterogeneity, did not allow identifying the issue of potential collinearity between the intercept and the exposure term. The true exposure surface ($X$) under Scenario A of \citet{Gryparis2009-mm} is not actually smooth because of the local heterogeneity term included in $X$.

Again, our simulation study reveals that it is highly important to consider the dependency information (spatial correlation) from the first-stage exposure model in the second-stage health analysis. 
Plugging in the point estimate, or completely ignoring the dependency information leads to the health effect estimator $\hat\beta_x$ with low quality (measured in terms of RMSE), and suffers from low coverage probability due to narrower credible intervals than they should have been. 
Our simulation result reveals that the proposed sparse MVN prior approach can be a good alternative to dense MVN prior or fully Bayesian approaches, especially when the number of participant locations is large or performing joint estimation of first and second models is infeasible.

\section{Real data analysis}
\label{sec:real}

We investigate associations between birth weight and traffic-related air pollution exposures, namely NO$_2$ exposures (a frequently used marker of traffic pollution) and contributions from gasoline sources (gasoline source-specific exposures), using the proposed approach as well as other state-of-the-practice approaches. Gasoline source-specific exposures were estimated by a previously developed Bayesian Spatial Multivariate Receptor Modeling (BSMRM) \citep{Park2018-sa} method while NO$_2$ exposures (pollutant-specific exposures) were estimated by the discrete process convolution model \citep{Higdon1998-pr,Higdon2002-gi}.  Appendix~\ref{appendix:B} contains the details of the exposure assessment and exposure surface plots over Harris County for selected days.  

We utilize TX DSHS vital statistics records for singleton live births for Harris County, TX, for the period of January 1-December 31, 2012. (This study was approved by the Institutional Review Board of Baylor College of Medicine and the Texas Department of State Health Services.) Using the geocoded coordinates of addresses at the time of delivery, we excluded duplicate locations. To account for potential confounding, the following covariates were included in the health effects model: fetal sex, maternal age, maternal race/ethnicity, maternal education, smoking during pregnancy, body mass index (BMI), hypertensive disorders of pregnancy, prenatal care, and census block group median household income. Other covariates, such as the presence of chronic diabetes, gestational diabetes, chronic hypertension, previous preterm birth, and other previous poor pregnancy outcomes were also considered initially but later excluded from the health model due to the lack of variation in the data. Following exclusions for duplicate or missing data, the number of births retained in the final dataset was 38,809.

\begin{figure}
    \centering
    \includegraphics[width=0.68\textwidth]{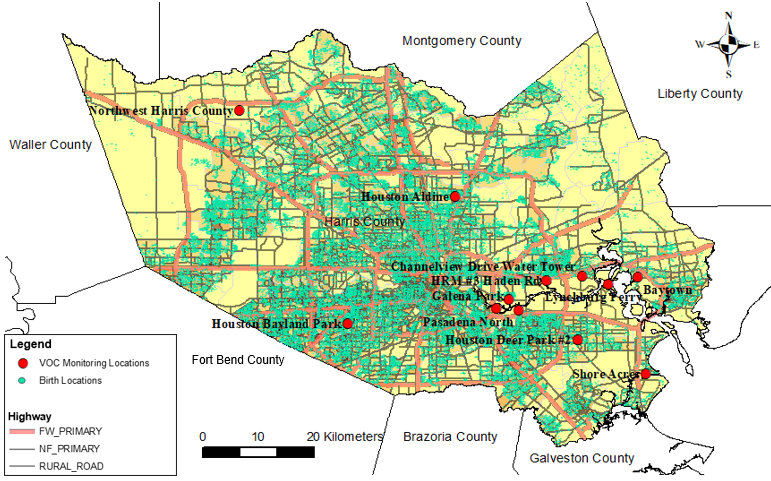}
    \caption{Map of 12 VOC monitoring stations (red) and approximate residential locations (green) of the study participants.}
    \label{fig:birthdata}
\end{figure}

Figure~\ref{fig:birthdata} shows the approximate residential locations of the study participants, as well as 12 monitoring stations measuring Volatile Organic Compounds (VOCs), used for estimating gasoline source-specific exposures. (The locations of NO$_2$ monitoring stations are shown in Figure~\ref{fig:logno22012} of Appendix~\ref{appendix:no2detail}). We estimated average exposures over the entire pregnancy period for each participant as the exposure metric. The average NO$_2$ exposures are based on daily averages of hourly measurements and the average gasoline source-specific exposures are based on daily measurements collected every six days. See Appendix~\ref{appendix:timeaveragedetail} for further details on handling time-averaged exposures and associated uncertainty for MVN prior approaches.  

With this large number of participants, the implementation of the dense MVN prior approach is not possible. As a matter of fact, Bayesian approaches (whether two-stage Bayesian or fully Bayesian) accounting for spatial exposure measurement errors have never been applied to real health outcome data of the size of this magnitude due to computational infeasibility. We performed the two-stage Bayesian analysis accounting for spatial exposure measurement error using the proposed sparse MVN prior approach (with 10 nearest neighbor conditioning sets and coordinate-based ordering) and compared the results with those from the analysis using the plug-in approach and the independent normal prior approach. 
For the plug-in approach, the mean of posterior predictive samples of exposures from the first stage analysis is used for the exposure (as if they were true values, i.e., without accounting for associated uncertainty) in the second stage analysis. We conducted both frequentist analysis and Bayesian analysis of the health model using the plug-in estimates of exposures as an independent variable. The independent normal prior approach uses the posterior predictive mean and the posterior predictive variance at each participant location from the first stage for the prior distribution for exposures in the second stage. The sparse MVN prior approach incorporates uncertainty and spatial correlation in predicted exposures across different participant locations into estimation by utilizing the second-order moment information in the posterior predictive distribution of exposures from the first stage. For Bayesian approaches, we used independent diffuse priors for all model parameters.

\begin{table}
  \footnotesize
  \centering
  \caption{Results examining associations between $\ln(\mathrm{NO}_2)$ exposure (ppbv) and birth weight (g) for live births in Harris county for 2012 from four regression models (simple plug-in regression and three different Bayesian uncertainty propagation methods). %
  We report scaled regression coefficients corresponding to approximately 0.69 unit change in $\ln(\mathrm{NO}_2)$ exposure, a doubling of $\mathrm{NO}_2$ concentration on the original scale.
  }
  \label{table:no2conti}
  \vspace{1mm}
  \begin{tabularx}{\textwidth}{X Y Y Y}
    \toprule
    Method & Estimate & SE/PSD & 95\% CI\\
     \midrule
 Plug-in (Non-Bayesian)&  -22.07 & 7.07 & (-35.92, -8.22) \\
 Plug-in  & -21.95 & 6.98 & (-35.97, -8.67) \\
 Independent normal  & -22.04 & 7.25 & (-35.75, -7.14)\\ 
 Sparse MVN & -21.97 & 7.47 & (-37.16, -8.12)\\  
     \bottomrule
  \end{tabularx}
  \vspace{-1mm}
\begin{flushleft}  
  Note: The following covariates are included in the health model: Fetal sex, Maternal age, Maternal race/ethnicity (non-Hispanic White, non-Hispanic Black, Hispanic/Latinx, Other/Unknown), Maternal education (High school or less, Some college, College or beyond), Smoking during pregnancy, BMI ($< 18.5$, $18.5$ to 24.9, 25 to 29.9, 30 to 34.9, $\ge 35$ kg/m$^2$), Hypertensive disorders of pregnancy, Prenatal care, Census block group median household income.
\end{flushleft}
\end{table}
 
Tables \ref{table:no2conti} and \ref{table:gasolineconti} present the analysis results for estimating the health effects of NO$_2$ exposures and gasoline source-specific exposures on birth weight, respectively. The health effects estimates (the coefficient estimates for the exposure variables) from four different analyses (frequentist analysis with Plug-in estimate and Bayesian analyses with Plug-in estimate, Independent normal prior, and Sparse MVN prior) along with the estimated coefficients for other covariates based on the Sparse MVN fit are provided. Uncertainty estimates, given as standard error (SE) or posterior standard deviation (PSD) and 95\% CIs (confidence intervals or credible intervals), are also provided next to the coefficient estimates. All four analyses suggest statistically significant negative associations (indicating adverse health effects) between NO$_2$ exposures and birth weight (Table \ref{table:no2conti}) or between gasoline source-specific exposures and birth weight (Table \ref{table:gasolineconti}) that are similar in magnitude and the uncertainty estimates follow similar patterns in the width of the credible intervals observed in the simulation studies. 
Compared to the plug-in approaches or the independent normal prior approach, the sparse MVN prior approach leads to larger uncertainty estimates for the health effect parameter ($\beta_x$), which is a natural consequence of accounting for uncertainty and spatial correlation in the exposure estimates and is expected to reflect the true uncertainty associated with the estimate of $\beta_x$. It needs to be noted that the spatial modeling of exposures in the first stage allows the estimates from the plug-in approaches and the independent approach to account for at least some exposure measurement errors (resulting from spatial misalignment error although not the uncertainty in parameter estimates of prediction models and correlation of exposure estimates), and consequently, the bias associated with those estimates does not seem to be too big, as also observed by \citet{Gryparis2009-mm}.
As shown in Table \ref{table:no2conti} from the sparse MVN prior approach (-21.97) controlling for other covariates in the model, a doubling of NO$_2$ exposures is associated with about 22g decrease in birth weight. Likewise, the health effect estimate of gasoline source-specific exposures from the sparse MVN prior approach (-21.80) in Table \ref{table:gasolineconti} can be interpreted as about 22g decrease in birth weight associated with an IQR increase (1.86 ppbC) of gasoline exposures controlling for other covariates in the model.

\begin{table}
  \footnotesize
  \centering
  \caption{Results examining associations between gasoline exposure (ppbC) and birth weight (g) for live births in Harris County for 2012 from four regression models (simple plug-in regression and three different Bayesian uncertainty propagation methods). 
  We report scaled regression coefficients corresponding to an interquartile range change (IQR = 1.86) in gasoline exposure.}
  \label{table:gasolineconti}
  \vspace{1mm}
 \begin{tabularx}{\textwidth}{X Y Y Y}
    \toprule
    Method &  Estimate & SE/PSD & 95\% CI\\
     \midrule 
  Plug-in (Non-Bayesian)  & -18.46 & 3.63  & (-25.57, -11.35) \\
 Plug-in   & -18.40 & 3.61 & (-25.33, -11.43) \\
 Independent normal & -18.59 & 3.47 & (-25.28, -11.71) \\ 
 Sparse MVN   & -21.80 & 4.38 & (-29.94, -13.11) \\
      \bottomrule
  \end{tabularx}
  \vspace{-1mm}
\begin{flushleft}  
  Note: The following covariates are included in the health model: Fetal sex, Maternal age, Maternal race/ethnicity (non-Hispanic White, non-Hispanic Black, Hispanic/Latinx, Other/Unknown), Maternal education (High school or less, Some college, College or beyond), Smoking during pregnancy, BMI ($< 18.5$, $18.5$ to 24.9, 25 to 29.9, 30 to 34.9, $\ge 35$ kg/m$^2$), Hypertensive disorders of pregnancy, Prenatal care, Census block group median household income.
\end{flushleft}
\end{table}

The results of running logistic regression models for low birth weight outcomes are presented in Table \ref{table:no2binary} and Table \ref{table:gasolinebinary} of Appendix~\ref{appendix:addtable_real}. Analyses of the effects of NO$_2$ exposures and gasoline source-specific exposures both indicated increased odds of an infant having low birth weight although the effects of NO$_2$ exposures were not statistically significant. This outcome may not be surprising because binary health outcomes contain inherently less information compared to continuous health outcomes as also noted in \citet{Gryparis2009-mm}. The effects of gasoline source-specific exposures were statistically significant, however. As shown in Table  \ref{table:gasolinebinary}, the odds ratio estimate from the sparse MVN prior approach ($\exp(0.14)=1.15$) indicates a 15\% increase in the odds of having an infant with low birth weight with an IQR increase (1.86 ppbC) of gasoline exposure. As in the case for birth weight modeled as a continuous variable, the odds ratio estimates obtained from the other approaches were slightly smaller (by about 1\%).

\section{Discussion}
While a two-stage Bayesian approach has been advocated as a method to propagate exposure uncertainty into health effect estimation in air pollution epidemiology research, the issue of scalability of the second-stage model fitting of Bayesian analysis has long been ignored. In fact, there has yet to be any practically implementable approach accounting for spatial correlations in exposure estimates when the number of participant locations is large. We present a scalable two-stage Bayesian approach utilizing a sparse multivariate normal prior based on a Vecchia approximation in the second-stage model fitting. Our simulation study revealed that the proposed sparse MVN prior approach provides significantly less biased health effect estimates and better quantifies the true uncertainty associated with the estimates than other (scalable) approaches, such as the plug-in approach which is most widely used in air pollution epidemiology studies. 
Also, comparison with the independent normal prior approach reveals that reflecting only marginal variance in exposure estimates is not sufficient, and information about spatial dependence must be taken into account.
Our data analysis on birth outcomes in Harris County, involving a sample size of $38,809$, also demonstrates that the uncertainty estimates of the health effects parameters of air pollution (both NO$_2$-specific and gasoline source-specific air pollution) estimated by the sparse MVN prior approach better account for exposure measurement errors from uncertainty in the exposure estimates. This is particularly advantageous because the dense MVN prior approach would be impractical for conducting the analysis in such cases with large sample sizes. 
We developed the R codes to facilitate the implementation of the proposed \textit{Sparse MVN} prior approach and other approaches discussed in this article. The R package \texttt{bspme} implementing the sparse MVN prior approach can be found at \url{https://github.com/changwoo-lee/bspme}, and we hope this package will help practitioners in air pollution epidemiology research.

A major advantage of the proposed two-stage Bayesian approach is that it can be applied to health effects analysis of any predicted/estimated exposures, including source-specific exposures (concerning the effects of multiple pollutants simultaneously), not just pollutant-specific exposures (concerning the effect of a single pollutant), as long as the exposure estimates along with their uncertainty information are available. Although source-specific exposures themselves represent exposures from multiple air pollutants, an important extension of our work would be to jointly estimate the health effects of multiple exposures (including multiple source-specific exposures) while accommodating cross-correlations across different exposures as well as spatial correlation and uncertainty in exposure estimates. 

Assessing the robustness of the two-stage Bayesian approaches to model misspecification is another avenue of future research. When the exposure model is misspecified, its effect on the accuracy of a health effect estimate is generally hard to analyze. 
There is growing literature on the modularized Bayesian approach \citep{Bayarri2009-ek, Plummer2015-bo, Jacob2017-qr} by cutting feedback from one stage to another to address the model misspecification problem. This leads to a posterior distribution of a second-stage parameter, called ``cut posterior'', which is both conceptually and mathematically different from a posterior distribution based on fully Bayesian and two-stage Bayesian approaches. 
The existing uncertainty propagation method called \textit{exposure simulation} or \textit{multiple imputation} methods \citep{Gryparis2009-mm, Comess2024-ot} aims to target the ``cut posterior'', although its connection with the modularized Bayesian approach is rarely discussed in the literature.
We emphasize that the main goal of the proposed sparse MVN prior approach is to provide a scalable alternative to the standard two-stage Bayesian approach (dense MVN prior approach) as well as to the fully Bayesian approach.
We believe that a comparison with the modularized Bayesian approach in the context of model misspecification problems is worth another investigation.

\section*{Software}

The R package \texttt{bspme} implementing the sparse MVN prior approach proposed in this article is available at
\url{https://github.com/changwoo-lee/bspme}.
\section*{Acknowledgements}

Conflict of Interest: None to declare.

\section*{Funding}

This research was supported by the National Institute of Environmental Health Sciences (NIEHS) of the National Institutes of Health (NIH) under R01ES031990.

\bibliography{paperpile.bib}
\bibliographystyle{apalike}

\clearpage 

\appendix

\counterwithin{figure}{section}
\counterwithin{table}{section}

\section{Appendix A}

\subsection{Details on simulation settings}
\label{appendix:detailsettings}
We first provide details on the exposure data generation scheme from the DPC model. We set $L=25$ gridpoints $\{u_1,\dots,u_{25}\} = \{0.2, 0.6, 1, 1.4, 1.8\}\times \{0.2, 0.6, 1, 1.4, 1.8\}$, regularly distributed over the domain $\calD = [0,2]\times [0,2]$. The true discrete process on the gridpoints, denoted as $G(x,y)$ for $(x,y)\in \{u_1,\dots,u_{25}\}$, is defined as  
\[
G(x,y) = \begin{cases}
3, & (x,y) = (0.2,1)\\
2,& (x,y) = (0.2,0.2)\text{ or }(1,0.2)\text{ or } (0.2,1.8)\\
1, & (x,y) = (1.8,0.6)\text{ or }(1,1)\text{ or }(1.8, 1.4)\text{ or }(1, 1.8)\\
0, & \text{otherwise}
\end{cases}
\]
for scenario A, and
\[
G(x,y) = \begin{cases}
3, & (x,y) = (1,1)\\
2,& (x,y) = (1.4, 0.2) \text{ or }(0.6,0.6)\text{ or } (0.2,1.8)\\
1, & (x,y) = (1, 0.2)\text{ or }( 1.8, 0.2 )\text{ or }(0.2, 1)\text{ or }(1.4, 1.4) \text{ or }(0.6, 1.8)\\
-1, & (x,y) =  (0.2, 0.2) \text{ or }(1.8, 1) \text{ or }(1.4, 1.8)\\
0, & \text{otherwise}
\end{cases}
\]
for scenario B. Based on those discrete processes, we generated true exposure at measurement locations and participant locations $(\bfX,\bfX^*)$ jointly, from the DPC model \eqref{eq:dpc} with $\mu = 3$ and bivariate Gaussian convolution kernel $K(s-u_l) = (2\pi\sigma_k^2)^{-1}\exp\left(-\|s-u_l\|_2^2/(2\sigma_k^2)\right)$ with standard deviation $\sigma_k = 0.4$. 
We set $\sigma_W = 0.1$ for the error standard deviation of the exposure measurement $\bfW$. Health data were generated with a (generalized) linear model \eqref{eq:contioutcome} and \eqref{eq:binaryoutcome}, with three different numbers of participant locations $n_y\in \{1000,2000,5000\}$, uniformly distributed over the domain. A total of 400 simulated data are considered. 

Next, we describe prior distributions considered in the first and second-stage model. For the first-stage DPC model parameters in \eqref{eq:dpc}, we consider independent prior distributions $\mu\sim \mathrm{N}(\bar{W},10^2)$ where $\bar{W} = n_w^{-1}\sum_{i=1}^{n_w}W(s_i)$, $\sigma^2_W\sim\mathrm{IG}(0.01, 0.01)$, and $\sigma^2_G\sim \mathrm{IG}(0.01, 0.01)$ where $\mathrm{IG}$ denotes inverse gamma distribution. For continuous outcomes, we consider a conjugate normal-inverse-gamma prior for the second-stage model parameters, namely $\sigma^2_Y\sim \mathrm{IG}(0.01, 0.01)$ and $(\beta_0, \beta_x, \beta_z)\given\sigma^2_Y \sim \mathrm{N}_3(\bm{0}, 100\sigma^2_Y\bfI)$, so that when $\bfX^*$ is fixed, the posterior is also a normal-inverse-gamma distribution, which enables to draw independent samples from the posterior. 
For binary outcomes, we consider normal prior $(\beta_0, \beta_x, \beta_z) \sim \mathrm{N}_3(\bm{0}, 100\bfI)$.

\subsection{Details on posterior inference procedure}
\label{appendix:postinferencedetail}

Posterior inferences were carried out with Gibbs samplers. In the first stage, after 10,000 burn-in iteration, we collected 1,000 posterior predictive samples with thinning interval of 5. 
In the second stage, after 10,000 burn-in iteration, we collected 2,000 samples with thinning interval of 5. 
When comparing second stage fitting times in Tables \ref{table:conti1000}, \ref{table:binary5000}, and Tables \ref{table:conti20005000}, \ref{table:binary10002000}, we used the wall-clock time taken for total 20,000 iterations.
All computations were performed on Intel E5-2697 v2 CPU with 128GB of memory. 

\subsubsection{First stage exposure model with discrete process convolution (DPC)}
\label{appendixsubsub:postdpc}
Let prior distributions be $\mu\sim \mathrm{N}(m_\mu, \sigma_\mu^2)$, $\bfG \given \sigma_G^2 \sim \mathrm{N}_{L}(\bm{0}, \sigma_G^2\bfI_L)$, $\sigma_G^2\sim \mathrm{IG}(a_G, b_G)$ and $\sigma_W^2\sim \mathrm{IG}(a_W, b_W)$. Denoting $\tilde\bfK = [\bm{1}_{n_w}, \bfK] \in \bbR^{n_w\times (1+L)}$, the Gibbs sampler is described in Algorithm~\ref{alg:firstdpc}.

\begin{algorithm}
\small
\caption{First stage Gibbs sampler with DPC model}\label{alg:firstdpc}
Initialize parameters and repeat: \\
\textbf{Step 1}: Sample $[\mu, \bfG] \given \bfW, \sigma_G^2, \sigma_W^2 \sim \mathrm{N}_{1+L}(\bfQ_1^{-1}\bfb_1, \bfQ_1^{-1})$ where
    \[
    \bfQ_1 = \sigma_W^{-2}\tilde\bfK^\top\tilde\bfK+\diag(\sigma_\mu^{-2},\sigma_G^{-2}\bfI_L), \quad \bfb_1 = [m_\mu/\sigma_\mu^2, \bm{0}_{L}]^\top + \sigma_W^{-2}\tilde\bfK^\top \bfW.
    \]\\
\textbf{Step 2}: Sample $\sigma_G^2 \given \bfW, \bfG \sim \mathrm{IG}(a_G+ L/2, b_G + \bfG^{\top}\bfG/2)$.\\
\textbf{Step 3}: Sample $\sigma_W^2 \given \bfW, \mu, \bfG \sim  \mathrm{IG}(a_W+ n_w/2, b_W + (\bfW - \bfK\bfG - \mu \bm{1}_{n_w})^{\top} (\bfW - \bfK\bfG - \mu \bm{1}_{n_w})/2)$.
\end{algorithm}

\subsubsection{Second stage health model with continuous outcome}
\label{appendixsubsub:postconti}
The second stage health model with continuous outcome can be written in a matrix form
\[
\bfY^* = \beta_0\bm{1}_{n_y} + \beta_x \bfX^* + \bfZ^*\bm\beta_z + \bm\epsilon, \quad \bm\epsilon\sim \mathrm{N}_{n_y}(0,\sigma_Y^2\bfI_{n_y}),
\]
where $\bfZ^*\in\bbR^{n_y\times p}$ is a matrix of $p$ covariates. Letting $\bm{\Phi} = [\bf1, \bfX^*,\bfZ^*]$ be a design matrix and $\bm\beta = [\beta_0,\beta_x,\bm\beta_z^\top]^\top$ be a coefficient, we can write compactly as $
\bfY^* = \bm{\Phi}\bm\beta + \bm\epsilon$.
We use the conjugate normal-inverse-gamma prior distribution $\sigma_Y^2\sim \mathrm{IG}(a_Y, b_Y)$ and $\bm\beta \given \sigma_Y^2 \sim \mathrm{N}_{2+p}(\bm{0}, \sigma_Y^2\bm\Sigma_\beta)$, so that the posterior distribution $\bm\beta, \sigma_Y^2 \given \bfY^*$ is also a normal-inverse-gamma distribution. 
With prior $\bfX^*\sim \mathrm{N}_{n_y}(\bm\mu_0, \bm\Sigma_0)$, the correponding Gibbs sampler is described in Algorithm~\ref{alg:second_conti}. Here, $\hat{\bm\beta} = (\bm\Sigma_\beta^{-1}+\bm{\Phi}^\top\bm{\Phi})^{-1}\bm{\Phi}^{\top}\bfY^*$, the posterior mean of $\bm\beta$. For the plug-in approach, step 1 is omitted.

\begin{algorithm}[h]
\small
\caption{Second stage Gibbs sampler for normal linear regression model}\label{alg:second_conti}
Initialize parameters and repeat: \\
\textbf{Step 1}: Sample $\bfX^* \given \bfY^*, \bm\beta, \sigma_Y^2 \sim \mathrm{N}_{n_y}\left(\bfQ_2^{-1}\bfb_2, \bfQ_2^{-1}\right)$ where 
\[
\bfQ_2 = \bm\Sigma_0^{-1} + (\beta_x^2/\sigma_Y^2)\bfI_{n_y}, \quad \bfb_2 = \bm\Sigma_0^{-1}\bm\mu_0 + (\beta_x/\sigma_Y^2)(\bfY^* - \beta_0\bm{1}_{n_y}-\bfZ^*\bm\beta_z)
\]\\
\textbf{Step 2}:  Sample $ \sigma_Y^2 \given \bfY^*, \bfX^* \sim \mathrm{IG}(a_Y + n_y/2, b_Y + (\bfY^{*\top}\bfY^* -
    \hat{\bm\beta}^\top (\bm\Sigma_\beta^{-1} + \bm{\Phi}^\top\bm{\Phi})\hat{\bm\beta}) /2  )$\\
\textbf{Step 3}: Sample $ \bm\beta \given \sigma_Y^2,\bfY^*, \bfX^* \sim \mathrm{N}_{2+p}\left( \hat{\bm\beta}, \sigma_Y^2(\bm\Sigma_\beta^{-1} + \bm{\Phi}^\top\bm{\Phi})^{-1}\right)$
\end{algorithm}

\subsubsection{Second stage health model with binary outcome}
\label{appendixsubsub:postbinary}

To fit the Bayesian logistic regression model with normal prior $\bm\beta \sim \mathrm{N}_{2+p}(\bm{0}, \bm\Sigma_\beta)$, we utilize the data augmentation strategy of \citet{Polson2013-gb}. That is, we introduce P\'olya-Gamma auxiliary variables $\bm\omega = (\omega_1,\dots,\omega_{n_y})$ which leads to conditionally conjugate update in the Gibbs sampler, described in Algorithm~\ref{alg:second_binary}. 
Here we denote $\bm\phi_i^\top$ be the $i$th row of the design matrix $\bm{\Phi}$. For the plug-in approach where $\bfX^*$ is fixed, step 1 is omitted.

\begin{algorithm}[h]
\small
\caption{Second stage Gibbs sampler for logistic regression model}\label{alg:second_binary}
Initialize parameters and repeat: \\
\textbf{Step 1}: Sample $\bfX^* \given \bfY^*, \bm\beta \sim \mathrm{N}_{n_y}\left(\bfQ_3^{-1}\bfb_3, \bfQ_3^{-1}\right)$, where 
\[
\bfQ_3 = \bm\Sigma_0^{-1} + \beta_x^2\diag(\bm\omega), \quad \bfb_3 = \bm\Sigma_0^{-1}\bm\mu_0 + \beta_x\diag(\bm\omega)((\bfY^* - 0.5\bm{1}_{n_y})/\bm\omega  - \beta_0\bm{1}_{n_y}-\bfZ^*\bm\beta_z)
\]\\
\textbf{Step 2}:  Sample 
\[
\bm\beta \given \bfY^*, \bfX^* \sim \mathrm{N}_{2+p}((\bm{\Phi}^\top\diag(\bm\omega)\bm{\Phi} + \bm\Sigma_{\beta}^{-1})^{-1}\bm{\Phi}^\top(\bfY^*-0.5\bm{1}_{n_y}), (\bm{\Phi}^\top\diag(\bm\omega)\bm{\Phi} + \bm\Sigma_{\beta}^{-1})^{-1}).
\]\\
\textbf{Step 3}: Sample $\omega_i\given \bfX^*, \bm\beta \indsim \mathrm{PG}(1, \bm\phi_i^\top\bm\beta )$ for $i=1,\dots,n_y$.
\end{algorithm}

\subsubsection{Fully Bayesian model with continuous health outcome}
\label{appendixsubsub:postfullconti}
Under the DPC model, the first and second-stage models can be jointly written as 
\begin{align}
\bfW  &= \mu\bm{1}_{n_w} + \bfK\bfG + \bfe, \quad \bfe\sim \mathrm{N}_{n_w}(0,\sigma_W^2\bfI_{n_w}),\label{eq:fullBayesconti1}\\
\bfY^* &= \beta_0\bm{1}_{n_y} + \beta_x (\mu \bm{1}_{n_y} + \bfK^*\bfG) + \bfZ^*\bm\beta_z + \bm\epsilon, \quad \bm\epsilon\sim \mathrm{N}_{n_y}(0,\sigma_Y^2\bfI_{n_y}),
\label{eq:fullBayesconti2}
\end{align}
where $\bfX^*$ corresponds to $\mu \bm{1}_{n_w} + \bfK^*\bfG$. We have parameter $\sigma^2_W$ which only appears at the first stage, $\beta_0,\beta_x,\bm\beta_z,\sigma^2_Y$ which only appear at the second stage, $\mu$ and $\bfG$ which appear at both stages, and $\sigma_G^2$ which only depends on $\bfG$. Let prior distributions be $\mu\sim \mathrm{N}(m_\mu, \sigma_\mu^2)$, $\bfG \given \sigma_G^2 \sim \mathrm{N}_{L}(\bm{0}, \sigma_G^2\bfI_L)$, $\sigma_G^2\sim \mathrm{IG}(a_G, b_G)$, $\sigma_W^2\sim \mathrm{IG}(a_W, b_W)$, $\sigma_Y^2\sim \mathrm{IG}(a_Y, b_Y)$, and $\bm\beta \given \sigma_Y^2 \sim \mathrm{N}_{2+p}(\bm{0}, \sigma_Y^2\bm\Sigma_\beta)$ with $\bm\beta = [\beta_0,\beta_x,\bm\beta_z^\top]^\top$. By rewriting \eqref{eq:fullBayesconti1} and \eqref{eq:fullBayesconti2} as 
\[
\underbrace{\begin{bmatrix}
\bfW \\ (\bfY^* - \beta_0\bm{1}_{n_y} - \bfZ^*\bm\beta_z)/\beta_x 
\end{bmatrix}}_{=\breve\bfR } = 
\underbrace{\begin{bmatrix}
\bm{1}_{n_w} & \bfK \\
\bm{1}_{n_y} & \bfK^*
\end{bmatrix}}_{=\breve\bfK}
\begin{bmatrix}
\mu\\
\bfG
\end{bmatrix} + \bfE, \quad \bfE \sim \mathrm{N}_{n_w+n_y}\left(\bm{0},\bm\Sigma_E\right), 
\]
where $\bm\Sigma_E = \diag(\sigma_W^2\bm{1}_{n_w},(\sigma_Y^2/\beta_x^2)\bm{1}_{n_y} )$ and we introduce new notations $\breve{\bfR}$ and $\breve\bfK$ as above. Also denoting $\bm{\Phi} = [\bm{1}_{n_y}, \mu \bm{1}_{n_y} + \bfK^*\bfG,\bfZ^*]$, 
$\hat{\bm\beta} = (\bm\Sigma_\beta^{-1}+\bm{\Phi}^\top\bm{\Phi})^{-1}\bm{\Phi}^{\top}\bfY^*$, the corresponding Gibbs sampler is described in Algorithm~\ref{alg:fulldpc_conti}.

\begin{algorithm}[h]
\small
\caption{Fully Bayesian DPC model Gibbs sampler with continuous outcome}\label{alg:fulldpc_conti}
Initialize parameters and repeat: \\
\textbf{Step 1}: Sample $[\mu, \bfG] \given \bfW, \bfY^*, \sigma_W^2, \bm\beta, \sigma_Y^2 \sim \mathrm{N}_{1+L}(\bfQ_4^{-1}\bfb_4, \bfQ_4^{-1})$ where 
\[
\bfQ_4 = \breve\bfK^\top\bm\Sigma_E^{-1}\breve\bfK + \diag(\sigma_\mu^{-2}, \sigma_G^{-2}\bm{1}_{L}), \quad \bfb_4 = \breve\bfK^\top \bm\Sigma_E^{-1}\breve\bfR + [m_\mu/\sigma_\mu^2,\bm{0}_L]^\top
\]\\
\textbf{Step 2}: Sample $\sigma_G^2 \given \bfW, \bfG \sim \mathrm{IG}(a_G+ L/2, b_G + \bfG^{\top}\bfG/2)$.\\
\textbf{Step 3}: Sample $\sigma_W^2 \given \bfW, \mu, \bfG \sim  \mathrm{IG}(a_W+ n_w/2, b_W + (\bfW - \bfK\bfG - \mu \bm{1}_{n_w})^{\top} (\bfW - \bfK\bfG - \mu \bm{1}_{n_w})/2)$.\\
\textbf{Step 4}: Sample $ \sigma_Y^2 \given \bfY^*, \mu, \bfG \sim \mathrm{IG}(a_Y + n_y/2, b_Y + (\bfY^{*\top}\bfY^* -
    \hat{\bm\beta}^\top (\bm\Sigma_\beta^{-1} + \bm{\Phi}^\top\bm{\Phi})\hat{\bm\beta}) /2  )$.\\
\textbf{Step 5}: Sample $ \bm\beta \given \sigma_Y^2,\bfY^*, \mu,\bfG \sim \mathrm{N}_{2+p}\left( \hat{\bm\beta}, \sigma_Y^2(\bm\Sigma_\beta^{-1} + \bm{\Phi}^\top\bm{\Phi})^{-1}\right)$,
\end{algorithm}

\subsubsection{Fully Bayesian model with binary health outcome}
\label{appendixsubsub:postfullbinary}

The derivation is similar to the previous case, except that the auxiliary P\'olya-Gamma variables need to be updated as well. Let prior distributions be $\mu\sim \mathrm{N}(m_\mu, \sigma_\mu^2)$, $\bfG \given \sigma_G^2 \sim \mathrm{N}_{L}(\bm{0}, \sigma_G^2\bfI_L)$, $\sigma_G^2\sim \mathrm{IG}(a_G, b_G)$, $\sigma_W^2\sim \mathrm{IG}(a_W, b_W)$, $\bm\beta = (\beta_0, \beta_x, \bm\beta_z)\sim \mathrm{N}_{2+p}(\bm{0}, \bm\Sigma_\beta)$. 
We introduce new notation 
\[
\bfR(\bm\omega) = \begin{bmatrix}\bfW \\ ((\bfY^*-0.5\bm{1}_{n_y})/\bm\omega - \beta_0\bm{1}_{n_y} - \bfZ^*\bm\beta_z)/\beta_x \end{bmatrix}, \quad \bm\Sigma_E(\bm\omega) = \diag(\sigma_W^2\bm{1}_{n_w},(\beta_x^2\bm\omega)^{-1})
\]
and $\bm{\Phi} = [\bm{1}_{n_y}, \mu \bm{1}_{n_y} + \bfK^*\bfG,\bfZ^*]$. The corresponding Gibbs sampler is described in Algorithm~\ref{alg:fulldpc_binary}.

\begin{algorithm}[h]
\small
\caption{Fully Bayesian DPC model Gibbs sampler with binary outcome}\label{alg:fulldpc_binary}
Initialize parameters and repeat: \\
\textbf{Step 1}: Sample $[\mu, \bfG] \given \bfW, \bfY^*, \sigma_W^2, \bm\beta, \bm\omega \sim \mathrm{N}_{1+L}(\bfQ_5^{-1}\bfb_5, \bfQ_5^{-1})$ where 
\[
\bfQ_5 = \breve\bfK^\top\bm\Sigma_E(\bm\omega)^{-1}\breve\bfK + \diag(\sigma_\mu^{-2}, \sigma_G^{-2}\bm{1}_{L}), \quad \bfb_5 = \breve\bfK^\top \bm\Sigma_E(\bm\omega)^{-1}\bfR(\bm\omega) + [m_\mu/\sigma_\mu^2,\bm{0}_L]^\top
\]\\
\textbf{Step 2}: Sample $\sigma_G^2 \given \bfW, \bfG \sim \mathrm{IG}(a_G+ L/2, b_G + \bfG^{\top}\bfG/2)$.\\
\textbf{Step 3}: Sample $\sigma_W^2 \given \bfW, \mu, \bfG \sim  \mathrm{IG}(a_W+ n_w/2, b_W + (\bfW - \bfK\bfG - \mu \bm{1}_{n_w})^{\top} (\bfW - \bfK\bfG - \mu \bm{1}_{n_w})/2)$.\\
\textbf{Step 4}: Sample 
\[
\bm\beta \given \bfY^*, \mu, \bfG \sim \mathrm{N}_{2+p}((\bm{\Phi}^\top\diag(\bm\omega)\bm{\Phi} + \bm\Sigma_{\beta}^{-1})^{-1}(\bm{\Phi}^\top(\bfY^*-0.5\bm{1}_{n_y})), (\bm{\Phi}^\top\diag(\bm\omega)\bm{\Phi} + \bm\Sigma_{\beta}^{-1})^{-1})
\]
\textbf{Step 5}: Sample $\omega_i\given \mu, \bfG, \bm\beta \indsim \mathrm{PG}(1, \bm\phi_i^\top\bm\beta )$ for $i=1,\dots,n_y$.
\end{algorithm}

\subsection{Additional Tables}
\label{appendix:moretable}

In addition to Tables \ref{table:conti1000} and \ref{table:binary5000}, we report simulation results with different numbers of participants with residential locations $n_y$. 
The additional simulation results in Table \ref{table:conti20005000} and Table \ref{table:binary10002000} are consistent with the results in Tables \ref{table:conti1000} and \ref{table:binary5000}. The plug-in and independent normal prior approaches suffer from high RMSE and low coverage probability across all settings. Especially, as the length of the credible interval becomes narrower as $n_y$ increases, the discrepancy of the coverage probability of the plug-in and independent normal prior approaches becomes even larger, for example in Table \ref{table:conti20005000}, Scenario B with $n_y = 5000$, the coverage probability of plug-in approach is only 28\% compared to 95\%. It demonstrates that the uncertainty propagation method that accounts for spatial correlation of exposure estimates (smoothness of the exposure surface) becomes even more important when $n_y$ is large.

\begin{table}
  \small
  \centering
  \caption{Simulation results of a Bayesian linear regression model with a continuous outcome based on 400 replicates. Bias and RMSE are calculated with the posterior mean estimator $\hat\beta_x$. The $\bbE[\ell_{0.95}]$ and Coverage indicate the average length and empirical coverage of the 95\% credible interval. Time corresponds to the wall-clock time to fit the second stage in seconds.}
  \label{table:conti20005000}
  \vspace{3mm}
  \begin{tabular}{cc c c c c c c c}
    \toprule
    Continuous outcome & Method & Bias & RMSE & $\bbE[\ell_{0.95}]$ & Coverage & Time (s)\\
     \midrule
    \multirow{8}{*}{\makecell{Scenario A\\$n_y = 2000$}} & True exposure & -0.001 & 0.021 & 0.089 & 95.6\% & 7.7\\
 & Plug-in & 0.042 & 0.099 & 0.099 & 44.0\% & 7.9\\
 & Independent normal & 0.035 & 0.090 & 0.097 & 47.6\% & 73.8\\ 
 & Sparse MVN (3nn) & 0.002 & 0.057 & 0.178 & 88.8\% & 123.3\\ 
 & Sparse MVN (5nn) &-0.014 & 0.060 & 0.236 & 93.2\% & 165.6\\ 
 & Dense MVN &0.019 & 0.051 & 0.217 & 95.9\% & 6865.8\\ 
 & Fully Bayesian  & 0.012 & 0.046 & 0.192 & 95.4\% & -\\ 
     \midrule
    \multirow{8}{*}{\makecell{Scenario B\\$n_y = 2000$}} & True exposure &  0.000 & 0.012 & 0.049 & 96.0\% & 6.7\\
 & Plug-in & 0.015 & 0.051 & 0.054 & 44.5\% & 6.9\\
 & Independent normal & 0.013 & 0.043 & 0.052 & 50.2\% & 54.5\\ 
 & Sparse MVN (3nn) & 0.003 & 0.030 & 0.098 & 90.0\% & 106.9\\ 
 & Sparse MVN (5nn) & -0.003 & 0.031 & 0.123 & 95.2\% & 144.0\\ 
 & Dense MVN & 0.004 & 0.024 & 0.116 & 98.0\% & 5539.1\\ 
 & Fully Bayesian  &0.005 & 0.024 & 0.102 & 96.5\% & -\\ 
    \midrule
    \multirow{8}{*}{\makecell{Scenario A\\$n_y = 5000$}} & True exposure &  0.000 & 0.014 & 0.056 & 95.1\% & 10.0\\
 & Plug-in & 0.045 & 0.101 & 0.063 & 30.8\% & 10.4\\
 & Independent normal & 0.038 & 0.091 & 0.062 & 32.3\% & 152.8\\ 
 & Sparse MVN (3nn) & 0.017 & 0.058 & 0.109 & 66.0\% & 221.9\\ 
 & Sparse MVN (5nn) & -0.002 & 0.052 & 0.147 & 86.6\% & 315.9\\ 
 & Dense MVN & 0.024 & 0.051 & 0.195 & 94.2\% & 50575.7\\ 
 & Fully Bayesian  & 0.011 & 0.041 & 0.164 & 95.3\% & -\\ 
     \midrule
    \multirow{8}{*}{\makecell{Scenario B\\$n_y = 5000$}} & True exposure &  0.000 & 0.008 & 0.031 & 94.5\% & 9.2\\
 & Plug-in & 0.016 & 0.051 & 0.034 & 28.0\% & 9.5\\
 & Independent normal &0.013 & 0.043 & 0.033 & 32.8\% & 144.1\\ 
 & Sparse MVN (3nn) & 0.008 & 0.029 & 0.060 & 72.0\% & 244.3\\ 
 & Sparse MVN (5nn) &0.002 & 0.027 & 0.081 & 85.0\% & 315.5\\ 
 & Dense MVN &0.003 & 0.022 & 0.105 & 98.0\% & 50263.5\\ 
 & Fully Bayesian  &0.004 & 0.022 & 0.086 & 95.2\% &-\\ 
     \bottomrule
  \end{tabular}
\end{table}

\begin{table}
  \small
  \centering
  \caption{Simulation results of a Bayesian logistic regression model with a binary outcome based on 400 replicates. Bias and RMSE are calculated with the posterior mean estimator $\hat\beta_x$. The $\bbE[\ell_{0.95}]$ and Coverage indicate the average length and empirical coverage of the 95\% credible interval. Time corresponds to the wall-clock time to fit the second stage in seconds.}
  \label{table:binary10002000}
  \vspace{3mm}
  \begin{tabular}{cc c c c c c c c}
    \toprule
    Binary outcome & Method & Bias & RMSE & $\bbE[\ell_{0.95}]$ & Coverage & Time (s)\\
     \midrule
    \multirow{8}{*}{\makecell{Scenario A\\$n_y = 1000$}} & True exposure &  0.008 & 0.105 & 0.394 & 93.7\% & 27.2\\
 & Plug-in & 0.039 & 0.147 & 0.417 & 85.4\% & 27.3\\
 & Independent normal & 0.049 & 0.152 & 0.428 & 85.4\% & 70.8\\ 
 & Sparse MVN (3nn) & -0.006 & 0.132 & 0.516 & 96.0\% & 98.6\\ 
 & Sparse MVN (5nn) & 0.007 & 0.133 & 0.544 & 96.5\% & 122.1\\ 
 & Dense MVN &  0.033 & 0.131 & 0.488 & 93.7\% & 1141.3\\ 
 & Fully Bayesian & 0.033 & 0.131 & 0.489 & 94.0\% & -\\ 
     \midrule
    \multirow{8}{*}{\makecell{Scenario B\\$n_y = 1000$}} & True exposure &  0.008 & 0.076 & 0.286 & 93.8\% & 28.4\\
 & Plug-in & 0.012 & 0.101 & 0.289 & 86.8\% & 29.1\\
 & Independent normal & 0.034 & 0.107 & 0.304 & 88.8\% & 73.2\\ 
 & Sparse MVN (3nn) & 0.005 & 0.090 & 0.352 & 96.0\% & 106.0\\ 
 & Sparse MVN (5nn) & 0.010 & 0.087 & 0.362 & 98.2\% & 121.5\\ 
 & Dense MVN & 0.021 & 0.088 & 0.339 & 96.8\% & 1087.6\\ 
 & Fully Bayesian & 0.020 & 0.088 & 0.336 & 96.0\% & -\\ 
 \midrule
     \multirow{8}{*}{\makecell{Scenario A\\$n_y = 2000$}} & True exposure &0.004 & 0.069 & 0.279 & 96.5\% & 48.6\\
 & Plug-in & 0.037 & 0.116 & 0.295 & 81.5\% & 49.5\\
 & Independent normal &0.046 & 0.120 & 0.303 & 80.5\% & 112.3\\ 
 & Sparse MVN (3nn) &0.003 & 0.096 & 0.372 & 93.0\% & 164.8\\ 
 & Sparse MVN (5nn) & -0.006 & 0.098 & 0.424 & 96.2\% & 219.4\\ 
 & Dense MVN &0.024 & 0.093 & 0.377 & 94.8\% & 4490.9\\ 
 & Fully Bayesian &0.023 & 0.093 & 0.372 & 94.8\% & -\\ 
     \midrule
    \multirow{8}{*}{\makecell{Scenario B\\$n_y = 2000$}} & True exposure &  0.007 & 0.050 & 0.202 & 97.5\% & 46.7\\
 & Plug-in & 0.010 & 0.082 & 0.204 & 80.0\% & 47.7\\
 & Independent normal & 0.031 & 0.085 & 0.215 & 79.8\% & 110.9\\ 
 & Sparse MVN (3nn) &0.005 & 0.064 & 0.256 & 94.8\% & 158.8\\ 
 & Sparse MVN (5nn) & 0.010 & 0.064 & 0.271 & 97.5\% & 211.6\\ 
 & Dense MVN & 0.016 & 0.062 & 0.258 & 96.8\% & 4977.8\\ 
 & Fully Bayesian & 0.015 & 0.062 & 0.252 & 96.5\% & -\\ 
     \bottomrule
  \end{tabular}
\end{table}

\clearpage

\section{Appendix B}
\label{appendix:B}
\subsection{Details on summarizing time-averaged exposure}
\label{appendix:timeaveragedetail}
In the real data analysis on birth outcomes, the true exposure of an individual subject is defined as an \textit{average exposure} over the period of pregnancy, which differs by subject. Let $t=1,\dots,T$ be time indices that cover the pregnancy period. 
Denote $X(s_i^*,t)$ be a true exposure of subject $i$ on time $t$.
If the exposure period of the subject $i$ is $E_i\subset \{1,\dots,T\}$ with $e_i = |E_i|$, we define the average exposure of subject $i$ as
\begin{equation}
\bar{X}(s_i^*):= \frac{1}{e_i}\sum_{t\in E_i}X(s_i^*,t), \quad i=1,\dots,n_y.  
\label{eq:avgexposure}
\end{equation}
From the first stage exposure analysis, we have the posterior predictive distribution of $X(s_i^*,t)$ for each $i=1,\dots,n_y$ and $t=1,\dots,T$. For the second stage health model, the quantity of interest is average exposures of individual subjects \eqref{eq:avgexposure}. 

The corresponding MVN prior approach summarizes the dependency structure of the predictive distribution of $(\bar{X}(s_1^*),\dots, \bar{X}(s_{n_y}^*))$ using $n_y$-dimensional MVN. We introduce the following simplifying assumption that whenever $t\neq t'$, $X(s_i^*,t)$ and $X(s_j^*,t')$ are independent for any $i,j=1,\dots,n_y$. Under this temporal independence assumption, we have
\begin{align*}
    \cov(\bar{X}(s_i^*),\bar{X}(s_j^*)) &=  \cov\left(\frac{1}{e_i}\sum_{t\in E_i}X(s_i^*,t),\frac{1}{e_j}\sum_{t\in E_j}X(s_j^*,t)\right)\\
    &= \frac{1}{e_ie_j}\cov\left( \sum_{t\in E_i\cap E_j}\cov(X(s_i^*,t),X(s_j^*,t)) \right)
\end{align*}
Equivalently, the covariance matrix of $(\bar{X}(s_1^*),\dots, \bar{X}(s_{n_y}^*))$, denoted as $\bar{\bfS}$, can be expressed as
\begin{equation}
\bar{\bfS} = \diag(e_1^{-1},\dots,e_{n_y}^{-1})\left(\sum_{t=1}^T \bfS^{(t)}\odot \bfM^{(t)} \right) \diag(e_1^{-1},\dots,e_{n_y}^{-1})
\end{equation}
where $\bfS^{(t)}$ is an $n_y\times n_y$ sample covariance of $\{X(s_i^*,t)\}_{i=1}^{n_y}$ based on posterior predictive samples, $\odot$ denotes elementwise multiplication between two matrices with same size, and $\bfM^{(t)}$ is an $n_y\times n_y$ binary matrix, acting as a masking operator, defined as $M^{(t)}_{ij} = 1$ if $t\in E_i\cap E_j$ and 0 otherwise. 
Note that the temporal independence assumption is only introduced to summarize posterior predictive distributions at different times in a tractable manner for time-averaged exposures, and does not imply that first-stage exposure analyses are conducted independently across time. 

\newpage 

\subsection{Details on first stage NO$_2$ pollutant-specific exposure prediction}
\label{appendix:no2detail}
\begin{figure}[h]
    \centering
    \includegraphics[width=0.9\textwidth]{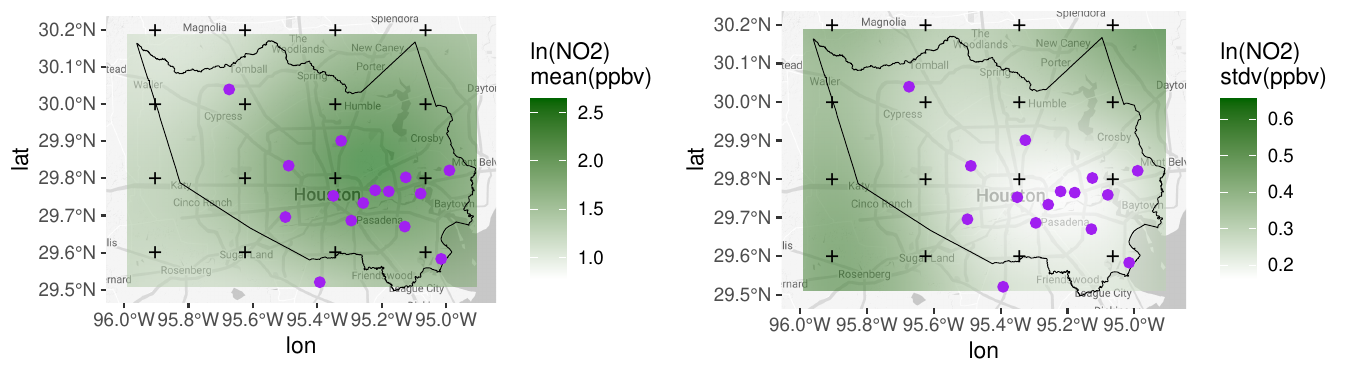}
    \caption{Posterior predictive mean and standard deviation of $\ln(\mathrm{NO}_2)$ exposure on Jan 10th, 2012 over the Harris County. Purple circles denote monitoring station locations and ``$+$'' denotes $L = 40$ discrete process grid locations.}
    \label{fig:logno22012}
\end{figure}

Here, we describe details of the first stage analysis for the NO$_2$ pollutant-specific exposure prediction used in the application to Harris County birth outcome data. We obtained hourly measurements of NO$_2$ concentrations from $n_w = 21$ monitoring stations from January 1, 2011 to December 31, 2012, which were subsequently aggregated to daily averages $(T=731)$. The natural logarithm of NO$_2$ daily averages was then used for $W$ in the exposure assessment to account for the nonnegativity of NO$_2$ concentrations.
Similar to the simulation studies, we adopt the discrete process convolution model with temporally varying intercepts, defined as
\begin{equation}
    W(s,t) = \mu(t) + \sum_{l=1}^L K(s-u_l)G(u_l,t) + \epsilon(s,t), \quad s\in\calD,\,\, t=1,\dots,T
    \label{eq:spatiotemporaldpc}
\end{equation}
where $\mu(t)$ is a temporally varying intercept term, $K(\cdot-u_l)$ is a Gaussian smoothing kernel centered at fixed grid locations $u_l\in\calD$ for $l = 1,\dots,L=40$. Here $G(u_l,t)\given \sigma_G^2 \iidsim \mathrm{N}(0, \sigma_G^2), l=1,\dots,L$, $t=1,\dots,T$ are normally distributed latent discrete process that is independent across space and time, and $\epsilon(s,t)$ is an independent normal error with mean zero and variance $\sigma_W^2$. Specifically, similar to \citet{Comess2024-ot}, we model temporally varying intercept as $\mu(t) = \sum_{j=1}^{14}\alpha_j\varphi_j(t)$, where $\varphi_j(t)$ is a B-spline basis functions defined over $[1,T]$ with 14 degrees of freedom. 
For the prior distributions, we used $\sigma_G^2\sim \mathrm{IG}(0.01, 0.01)$, $\sigma_W^2\sim \mathrm{IG}(0.01, 0.01)$, and spline basis coefficients $\bm\alpha\sim \mathrm{N}_{14}(\bm{0}, 100^2\bfI_{14})$. The posterior inference procedure is similar to Algorithm \ref{alg:firstdpc}, where few changes include an additional step to sample spline basis coefficients $\bm\alpha$ from its full conditional, which is normal as well, and sample variance parameters $\sigma_G^2$ and $\sigma_W^2$ based on $T\times L$ and $T\times n_w$ squared residual terms, respectively.
Figure~\ref{fig:logno22012} summarizes a posterior predictive distribution on a specific day obtained from the first-stage model.

\newpage

\subsection{Details on first stage gasoline source-specific exposure prediction}
\label{appendix:gasolinedetail}

Here, we describe details of the first stage analysis for the gasoline source-specific exposure prediction used in the application to Harris County birth outcome data in Section 5. We used the Bayesian spatial multivariate receptor model \citep[][BSMRM]{Park2018-sa} developed for source apportionment and prediction of spatially correlated source-specific exposures (source contributions). Unlike pollutant-specific exposure, source-specific exposures are not directly measured even at the monitoring locations but are assumed to be latent variables that explain variability in multiple pollutants contributed by individual source categories. We obtained volatile organic compounds (VOCs) data collected from $n_w = 12$ monitoring stations every 6th day during 2010-2012 ($T=183)$. We selected 17 important VOC species and used the same priors, hyperparameter settings, and identifiability conditions as those in the real analysis section of \cite{Park2018-sa}.

\begin{figure}[h]
    \centering
    \includegraphics[width=\textwidth]{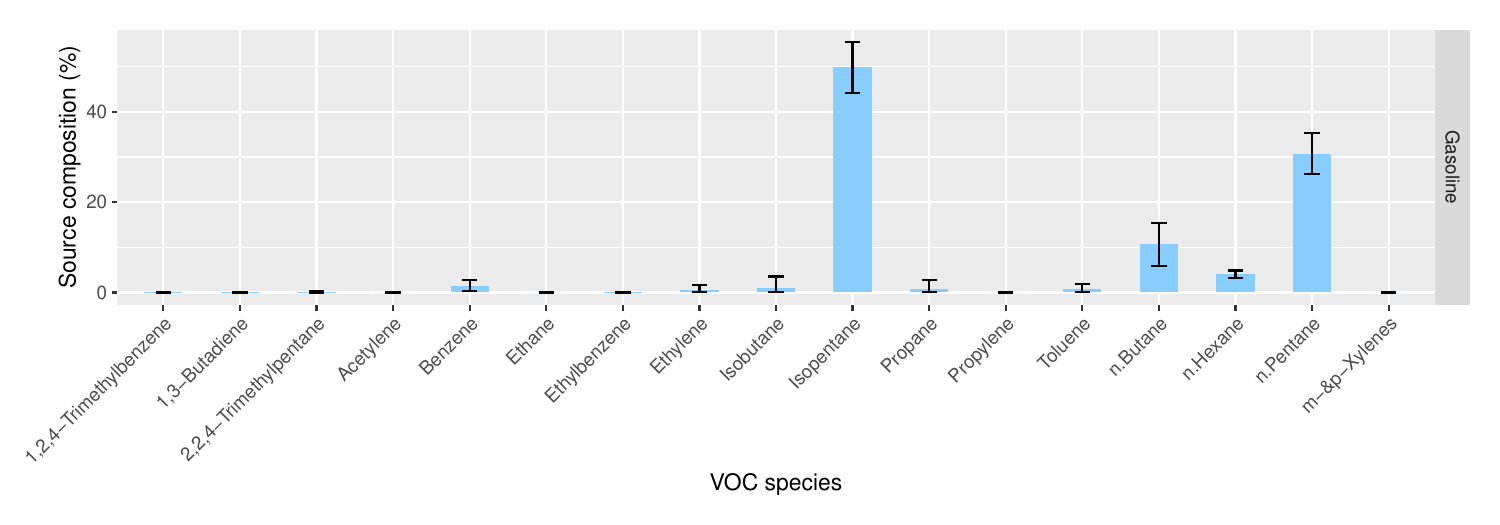}
    \caption{Estimated gasoline source composition profile with 95\% credible intervals.}
    \label{fig:Gasoline_compo}
\end{figure}

\begin{figure}
    \centering
    \includegraphics[width=\textwidth]{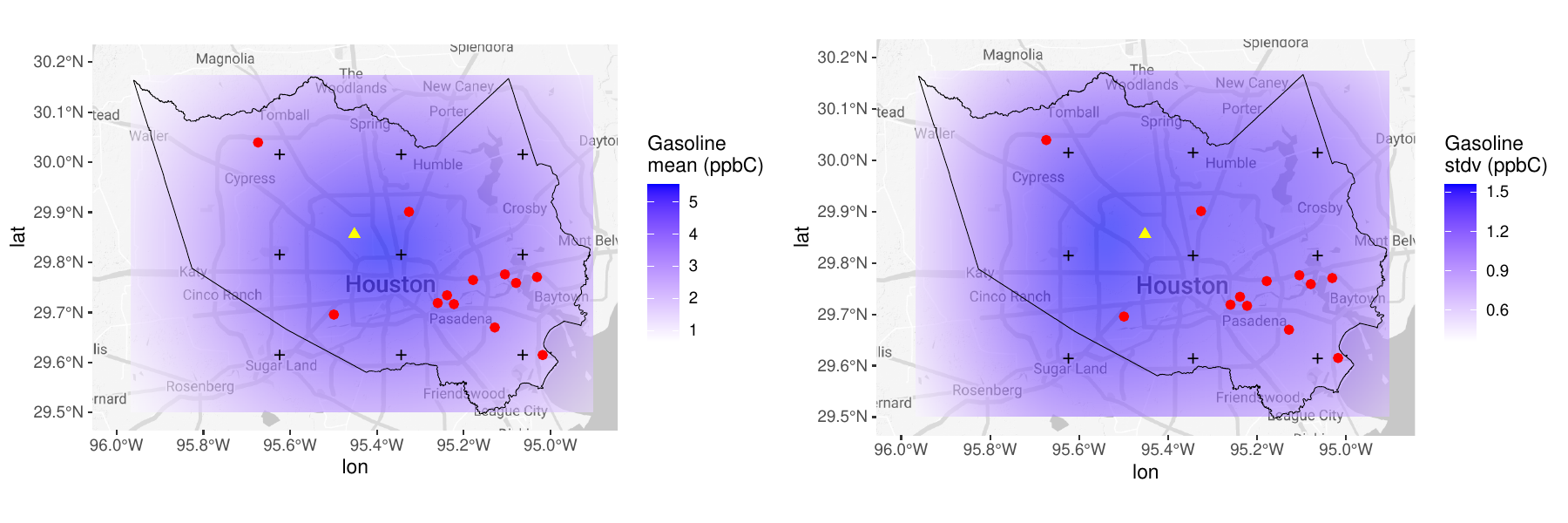}
    \caption{Posterior predictive mean and standard deviation of gasoline exposure on Jan 10th, 2012 over Harris County. Red circles denote $n_w=9$ monitoring station locations, and ``$+$'' denotes $L = 9$ discrete process grid locations. A yellow triangle denotes the unmonitored location for which time series plot of gasoline source-specific exposures is shown in Figure~\ref{fig:gasoline_unmonitored}.}
    \label{fig:gasoline2012}
\end{figure}

\begin{figure}
    \centering
    \includegraphics[width=\textwidth]{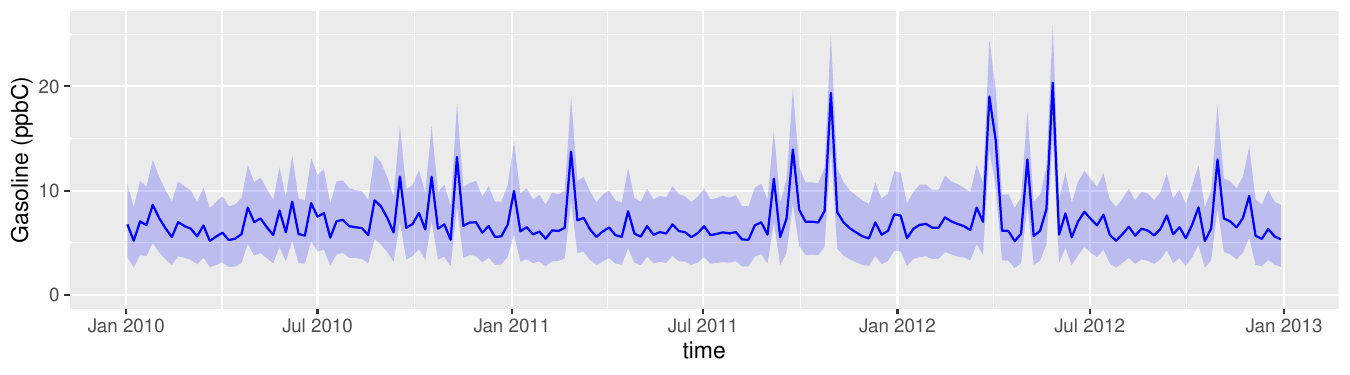}
    \caption{Predicted gasoline source-specific exposures during 2010-2012 along with 95\% credible intervals at an unmonitored site, marked as a triangle in Figure~\ref{fig:gasoline2012}. }
    \label{fig:gasoline_unmonitored}
\end{figure}

The estimated gasoline source composition profile, along with the uncertainty estimates (95\% credible intervals), is given in Figure~\ref{fig:Gasoline_compo}. 
Figures~\ref{fig:gasoline2012} and \ref{fig:gasoline_unmonitored} illustrate the predicted gasoline source contribution surface on a specific day (along with uncertainty estimates) and the time series of the predicted gasoline source contributions on a specific location, respectively.

\vspace{10mm}

\clearpage 

\subsection{Additional tables from the second stage real data analysis}
\label{appendix:addtable_real}

\begin{table}[h]
  \footnotesize
  \centering
  \caption{Results examining associations between $\ln(\mathrm{NO}_2)$ exposure (ppbv) and low birth weight (binary) for live births in Harris County for 2012 from four logistic regression models (simple plug-in regression and three different Bayesian uncertainty propagation methods). 
  We report scaled regression coefficients corresponding to approximately 0.69 unit change in $\ln(\mathrm{NO}_2)$ exposure, a doubling of $\mathrm{NO}_2$ concentration on the original scale.}
  \label{table:no2binary}
  \vspace{1mm}
  \begin{tabularx}{\textwidth}{X Y Y Y}
    \toprule
    Method & Estimate& SE/PSD & 95\% CI\\
     \midrule
 Plug-in (Non-Bayesian)  & 0.146 & 0.111 & (-0.071, 0.365) \\
 Plug-in   & 0.137 & 0.118 & (-0.096, 0.353)  \\
 Independent normal & 0.146  & 0.108  & (-0.064, 0.359) \\ 
 Sparse MVN  & 0.131  & 0.106 & (-0.080, 0.337) \\
     \bottomrule
  \end{tabularx}
  \vspace{-1mm}
\begin{flushleft}  
  Note: The following covariates are included in the health model: Fetal sex, Maternal age, Maternal race/ethnicity (non-Hispanic White, non-Hispanic Black, Hispanic/Latinx, Other/Unknown), Maternal education (High school or less, Some college, College or beyond), Smoking during pregnancy, BMI ($< 18.5$, $18.5$ to 24.9, 25 to 29.9, 30 to 34.9, $\ge 35$ kg/m$^2$), Hypertensive disorders of pregnancy, Prenatal care, Census block group median household income.
\end{flushleft}
\end{table}

\begin{table}[h]
  \footnotesize
  \centering
  \caption{Results examining associations between gasoline exposure (ppbC) and low birth weight (binary) for live births in Harris County for 2012 from four logistic regression models (simple plug-in regression and three different Bayesian uncertainty propagation methods). 
  We report scaled regression coefficients corresponding to an interquartile range change (IQR = 1.86) in gasoline exposure.}
  \label{table:gasolinebinary}
  \vspace{1mm}
  \begin{tabularx}{\textwidth}{X Y Y Y}
    \toprule
    Method  & Estimate & SE/PSD & 95\% CI\\
     \midrule
  Plug-in (Non-Bayesian)  & 0.136 & 0.059 & (0.021, 0.252)\\
 Plug-in  & 0.139 & 0.061 & (0.016, 0.254)\\
 Independent normal & 0.137 & 0.059 & (0.018, 0.252)\\ 
 Sparse MVN  & 0.143 & 0.060 & (0.026, 0.266)\\
     \bottomrule
  \end{tabularx}
  \vspace{-1mm}
\begin{flushleft}  
  Note: The following covariates are included in the health model: Fetal sex, Maternal age, Maternal race/ethnicity (non-Hispanic White, non-Hispanic Black, Hispanic/Latinx, Other/Unknown), Maternal education (High school or less, Some college, College or beyond), Smoking during pregnancy, BMI ($< 18.5$, $18.5$ to 24.9, 25 to 29.9, 30 to 34.9, $\ge 35$ kg/m$^2$), Hypertensive disorders of pregnancy, Prenatal care, Census block group median household income.
\end{flushleft}
\end{table}

\end{document}